\documentclass[%
preprint,
bibnotes,
 citeautoscript,
 amsmath,amssymb,
 longbibliography,
aip,
floatfix,
]{revtex4-2}

\usepackage{amsmath}
\usepackage{amsthm}
\usepackage{amsfonts}
\usepackage{amssymb}
\usepackage{graphicx}
\usepackage{numprint}
\usepackage{mathtools}
\usepackage[colorlinks=true,linkcolor=blue,citecolor=blue,urlcolor=cyan,bookmarks,breaklinks=true]{hyperref}
\usepackage{enumitem}
\usepackage{graphicx}
\usepackage{dcolumn}
\usepackage{bm}
\usepackage{xcolor}

\usepackage[caption=false]{subfig}
\usepackage[percent]{overpic}

\begin{document}
\title{Hole Statistics of Equilibrium 2D and 3D Hard-Sphere Crystals}

\begin{abstract}
The probability of finding a spherical ``hole'' of a given radius $r$  contains crucial structural information about many-body systems. 
Such hole statistics, including the void conditional nearest-neighbor probability functions $G_V(r)$, have been well studied for hard-sphere fluids in $d$-dimensional Euclidean space $\mathbb{R}^d$.
However, little is known about these functions for hard-sphere crystals for values of $r$ beyond the hard-sphere diameter, as large holes are extremely rare in crystal phases.
To overcome these computational challenges, we introduce a biased-sampling scheme that accurately determines hole statistics for equilibrium hard spheres on ranges of $r$ that far extend those that could be previously explored.
We discover that $G_V(r)$ in crystal and hexatic states exhibits oscillations whose amplitudes increase rapidly with the packing fraction, which stands in contrast to $G_V(r)$ for fluid states that is monotonic in $r$.
The oscillations in $G_V(r)$ for 2D crystals are strongly correlated with the local orientational order metric in the vicinity of the holes, and variations in $G_V(r)$ for 3D states indicate a transition between tetrahedral and octahedral holes, demonstrating the power of $G_V(r)$ as a probe of local coordination geometry.
To further study the statistics of interparticle spacing in hard-sphere systems, we compute the local packing fraction distribution $f(\phi_l)$ of Delaunay cells, and find that for $d\leq 3$, the excess kurtosis of $f(\phi_l)$ switches sign at a certain transitional global packing fraction. 
Our investigation facilitates the study of structural and bulk properties of materials that involve the creation of rare large holes, such as the solubility of alloys.
\end{abstract}
\author{Haina Wang}
\affiliation{\emph{Department of Chemistry, Princeton University}, Princeton, New Jersey, 08544, USA}
\author{David A. Huse}
\affiliation{\emph{Department of Physics, Princeton University}, Princeton, New Jersey, 08544, USA}
\author{Salvatore Torquato}
\email[]{Email: torquato@princeton.edu}
\affiliation{\emph{Department of Chemistry, Department of Physics, Princeton Institute of Materials, and Program in Applied and Computational Mathematics, Princeton University}, Princeton, New Jersey 08544, USA}
\date{\today}
\maketitle
\newpage

\section{Introduction}
\label{intro}


The venerable idealized hard-sphere model has been fruitfully employed to understand fundamentally the structure and bulk properties of a broad range of realistic systems with strong short-ranged repulsive interactions, including crystals and liquids of noble gases, glasses, granular media, and biological systems \cite{Ha86,To02a, To10c}.
The determination of the equilibrium phase diagrams for two-dimensional (2D) hard-disk and three-dimensional (3D) hard-sphere systems has been an active area of study since the pioneering studies of Alder and Wainwright \cite{Al57, Al62}, who presented numerical evidence for entropy-driven disorder-order phase transitions.
The 3D hard-sphere system undergoes a first-order  liquid-solid phase transition, with the corresponding freezing and melting packing fractions $\phi_f = 0.494$ and $\phi_m = 0.545$, respectively \cite{Ho68}. 
The densest packing fraction for the 3D hard-sphere crystal is given by $\phi_J = \pi/\sqrt{18}$, which is achieved by the  uncountably infinite stacking variants of hexagonal close-packed layers (e.g., fcc, hcp, etc.). 
There exist strong numerical evidence that the fcc lattice has the highest entropy among all possible stacking variants \cite{Bo97b, Mau99}.
For the equilibrium 2D hard-disk system, recent works \cite{Ma06, Ko08, Ber11} show that its phase diagram contains a fluid phase, a hexatic phase with short-range translational order and quasi-long-range orientational order, and a triangle-lattice crystal phase with densest possible packing fraction $\phi_J = \pi/\sqrt{12}$.
The freezing point for the hard-disk system, which corresponds to the transition between the fluid and the hexatic phases, occurs at $\phi_f = 0.69$ \cite{Ho68, Ko08}. 
A continuous phase transition between the hexatic and the crystal phases occurs at $\phi_h = 0.72$ \cite{Ber11}.
In this work, we focus on the ``hole'' statistics of 2D and 3D hard-sphere crystals.

A key microstructure descriptor used to probe the hole statistics in many-body systems is given by the ``void'' nearest-neighbor functions \cite{Re59, To90b, To02a}.
These functions, including the void probability density function $H_V(r)$ and the associated cumulative and conditional probability functions $E_V(r)$ and $G_V(r)$, have been extensively studied for hard spheres along the disordered (fluid) branch \cite{Re59, He60, To90b, To95a, To02a}.
The void nearest-neighbor conditional probability distribution function $G_V(r)$ is defined such that $\rho s_1(r)G_V(r)dr$ gives the probability of finding a particle at distance between $r$ and $r + dr$, given that one has found a hole of radius $r$ centered at a random point, where $s_1(r)$ is the surface area of a sphere with radius $r$ \cite{Re59, To02a}.
It is known that $G_V(r)$, $H_V(r)$ and $E_V(r)$ (defined in Sec. \ref{def}) can be expressed as integrals over all the $n$-particle correlation functions $g_2, g_3, \dots$ \cite{Re59, To02a}, and contain crucial information about equilibrium and nonequilibrium properties, such as the pressure and the excess chemical potential of equilibrium hard-sphere systems \cite{To90g, To02a}, as well as the degree to which a many-body system is out of equilibrium \cite{Wa23}.
Moreover, these functions play an important role in the theoretical study of liquids and amorphous matter \cite{To90b, To90c, Ri95b, To95a}, granular flows \cite{La13}, stellar dynamics \cite{Ch43}, disordered photonic materials \cite{Zh17a, Kl22}, as well as covering and quantizer problems in discrete geometry \cite{To10d}.
Because of these theoretical and practical interests associated with the nearest-neighbor functions, it is highly desirable to precisely determine them for many-body systems with standard interparticle interactions, including the hard-sphere model.

For one-dimensional (1D) equilibrium hard rods, $H_V(r), E_V(r)$ and $G_V(r)$ are known analytically at all packing fractions $\phi$ \cite{Sa53, To90b}. 
Exact solutions for these functions are not available for equilibrium hard spheres in two and higher dimensions, but can be very well approximated along the fluid (disordered) branch via scaled-particle theory due to Reiss et al. \cite{Re59}.
These authors showed that $G_V(r)$ for a hard-sphere fluid is a monotonically increasing function in $r$ and its asymptotic for large $r$ is given by
\begin{equation}
    G_V(r) = a_0 + \frac{a_1}{r} + \frac{a_2}{r^2} + \dots + \frac{a_{d-1}}{r^{d-1}} + \mathcal{O}\left(\frac{1}{r^d}\right), \quad r \gg \mathcal{D},
    \label{GvSeries}
\end{equation}
where $\mathcal{D}$ is the sphere diameter and $a_i$ are coefficients that depend on $\phi$.
Equation (\ref{GvSeries}) provides a good approximation at small and intermediate $r$ with $r \geq \mathcal{D}/2$ as well \cite{Re59, He60}.
Somewhat more accurate expressions for $a_i$ and thus $G_V(r)$ were obtained by Torquato et al \cite{To90b}.
Torquato further showed that the form (\ref{GvSeries}) is still valid in the the metastable extension of the liquid branch of hard-sphere systems \cite{To95a}.

Compared to disordered states, relatively less is known about the void nearest-neighbor functions for crystal states.
While theoretical and simulation studies have been performed to determine concentrations and free energies of vacancies in 2D and 3D equilibrium hard-sphere crystals \cite{Ba00b, Pr01, Ro12, Us17, Pu18}, they were mainly concerned with the void nearest-neighbor functions at the specific distance $r=\mathcal{D}$, which determines the thermodynamic properties of hard-sphere systems. 
Studying nearest-neighbor statistics in crystals for a large range of $r$ up to $r = 2\mathcal{D}$ is challenging, because large holes are extremely rare at high packing fractions.
The standard method of computing these functions, which involves sampling nearest-neighbor distances from particle centers to randomly placed test points, becomes inaccurate with greater than $100\%$ error when $r \sim 0.7 \mathcal{D}$ for 2D and 3D crystals near melting, at which $E_V(r) \sim 10^{-6}$ \cite{Wa23, Note5}.
\footnotetext[5]{Because $H_V(r)$ for hard-sphere systems decay asymptotically as $\exp(-r^d)$ \cite{To90g}, it is not possible to significantly improve upon this computational limit by simply scaling up computational resources.}
Furthermore, simulations in the grand canonical (GC) ensemble are required to efficiently study hole statistics, as the GC ensemble allows particles to be added and removed to create or migrate large holes.
However, standard Monte-Carlo (MC) simulations in the grand canonical ensemble are not feasible in the crystal phase, since particle additions result in a very high rejection rate.

In this work, we overcome these computational challenges by introducing a biased-sampling scheme for equilibrium hard spheres in the first three spatial dimensions that enables accurate and efficient sampling of extremely rare large holes.
Our algorithm is based on Monte Carlo simulations of hard spheres and a test point that repels its nearest hard-sphere neighbor via a biasing potential to create large holes. 
We apply our algorithm to compute void nearest-neighbor functions on ranges of $r$ beyond the hard-sphere diameter, up to rare holes that occur with probabilities at least 5 orders of magnitude smaller than the sensitivity limit of the aforementioned unbiased approach, which samples nearest-neighbor distances to random test points.
For 2D crystals, hexatic phases, and dense fluids with $\phi \leq 0.73$, we are able to compute $G_V(r)$ within $\pm 3\%$ errors up to $r = 1.95 \mathcal{D}$ and $\pm 5\%$ errors up to $r = 2 \mathcal{D}$, which is around 2.5 times larger than the ranges on which one can compute these functions accurately via the unbiased approach.
At these packing fractions, the sensitivity of hole sampling are improved by more than 40 orders of magnitude compared to the unbiased scheme, and holes with probability down to $E_V(r)\sim 10^{-48}$ can been detected.
For 2D crystals with $0.74 \leq \phi \leq 0.85$ and 3D crystals with $0.54 \leq \phi \leq 0.60$, we are able to accurately determine $G_V(r)$ up to at least $1.1 \mathcal{D}$ within $300$ hours on 200 parallel threads of 2.8 GHz Intel CPUs, which improves the computational time by at least 5 orders of magnitude compared to the unbiased method for all values of $\phi$ that we study.
This capability enables us to investigate the relationship between variations in the void nearest-neighbor functions and the local coordination geometry of large holes in crystals, which could not be done in previous works due to the limited ranges of the hole statistics.
To further characterize the distribution of interparticle spacing in hard-sphere systems, we also study the local packing fraction distribution $f(\phi_l)$ of Delaunay cells as a function of the global packing fraction in the first three spatial dimensions.  

We validate our methodology by comparing our numerical nearest-neighbor statistics for 1D hard rods with the known analytical expressions \cite{To02a}, and find that excellent agreement up to $r = 2\mathcal{D}$ can be achieved with a small system size of $\langle N\rangle = 200$.
Importantly, for 2D and 3D hard-sphere crystals, we find for the first time that that the conditional void probability function $G_V(r)$ exhibits oscillations whose amplitudes increase rapidly with the packing fraction. 
The function $G_V(r)$ is also oscillatory for 2D hard spheres in the hexatic phase.
On the other hand, $G_V(r)$ for disordered fluid states increases monotonically in $r$, even for high-density fluids near freezing.
We discover that oscillations in $G_V(r)$ for the 2D crystal are strongly correlated with the Nelson-Halperin orientational order metric $\psi_6$ \cite{Ne79} in the vicinity of the holes.
Specifically, the minima in $G_V(r)$ correspond to ``stable'' hole sizes that preserve the local hexatic order of the crystal, i.e., holes created by removing 1, 3, 7, \dots particles in the triangle lattice.
For 3D hard-sphere fcc crystals, we observe a shoulder in the first peak of $G_V(r)$, which indicates the transition between tetrahedral and octahedral holes. 
These findings demonstrate the power of $G_V(r)$ for probing the compatibility of the crystal structures with spherical cavities of given sizes, which is closely related to the local coordination geometry of such holes.
We find that in each of the first three space dimensions, the excess kurtosis of the local packing fraction distribution of Delaunay cells $f(\phi_l)$ increases with the global packing fraction $\phi$, and switches from negative to positive at a certain dimension-dependent transition packing fraction $\phi^*$.
The values of $\phi^*$ lie in the crystal phases for $d = 2$ and 3.
This switch from a platykurtic to a leptokurtic behavior in $f(\phi_l)$ as $\phi$ increases reflects a transition between weakly correlated displacements of hard particles at lower $\phi$ to highly correlated collective displacements at higher $\phi$.

In Sec. \ref{def}, we provide preliminary definitions and background.
In Sec. \ref{meth}, we describe our biased-sampling scheme to compute hole statistics of equilibrium systems.
Section \ref{res} presents results for the hole statistics of equilibrium hard spheres obtained via the sampling algorithm.
In Sec. \ref{sec_localphi}, we study the local packing distribution of Delaunay cells.
We provide conclusive remarks in Sec. \ref{conc}.

\section{Definitions and Preliminaries}
\label{def}
In this section, we introduce some fundamental concepts on void nearest-neighbor functions for classical many-body systems, as well as their relations to important thermodynamic properties.
We also provide exact expressions for the void nearest-neighbor functions for 1D hard-rod fluids \cite{To90c}, as well as accurate approximations of these functions for 2D hard-sphere systems along the disordered branch \cite{To95a}.

Consider a many-body system in $d$-dimensional Euclidean space $\mathbb{R}^d$. 
The void nearest-neighbor probability density function $H_V(r)$ is defined as \cite{To90b}
\begin{equation}
    \parbox{30em}{$H_V(r)dr =$ probability that at an arbitrary located point in the system, the nearest particle center lies at a distance between $r$ and $r + dr$.}
\end{equation}
The associated complementary cumulative distribution function, also called the void exclusion probability function $E_V(r)$, is given by \cite{To90b}
\begin{equation}
\begin{split}
    E_V(r) &= 1 - \int_0^r H_V(r')dr' \\
    &\parbox{30em}{= probability of finding a spherical region of radius $r$ centered at some arbitrary point void of particle centers.}
\end{split}
\label{def_ev}
\end{equation}
This definition is often given succinctly as the probability of finding a hole of radius $r$.
Finally, the associated conditional probability function $G_V(r)$ is defined as \cite{To90b}
\begin{equation}
    G_V(r) = \frac{H_V(r)}{\rho s_1(r) E_V(r)} = \frac{-1}{\rho s_1(r)}\frac{d \ln E_V(r)}{d r},
    \label{def_gv}
\end{equation}
where $s_1(r) = d \pi^{d/2}r^{d-1}/\Gamma(1+d/2)$ is the surface area of a $d$-dimensional sphere of radius $r$.
The function $G_V(r)$ has the interpretation that $\rho s_1(r) G_V(r) dr$ is the conditional probability that, given a spherical cavity of radius $r$ empty of particle centers, there exist particle centers in the spherical shell of volume $s_1(r)dr$ encompassing the cavity.

The void nearest-neighbor functions are related to crucial thermodynamic quantities. 
For example, the reduced pressure of an equilibrium monodisperse hard-sphere system at packing fraction $\phi$ is given by \cite{To90c, To02a}
\begin{equation}
    \frac{p}{\rho k_B T} = G_V(\infty) = 1 + 2^{d-1}\phi G_V(\mathcal{D}),
    \label{pressure}
\end{equation}
where $\rho = \phi/v_1(\mathcal{D}/2)$ is the number density, $v_1(r) = \pi^{d/2} r^d/\Gamma(1+d/2)$ is the volume of a $d$-dimensional sphere of radius $r$, and $T$ is the temperature.
Furthermore, the excess chemical potential $\mu'$ of an equilibrium hard-sphere system is given by  \cite{To90g, To02a}
\begin{equation}
     \mu' = \mu - \mu_{\text{id}} = - k_B T\ln[E_V(\mathcal{D})],
     \label{muprime_ev}
\end{equation}
where $\mu_{\text{id}} = k_B T \ln (\rho \Lambda^d)$ is the chemical potential of the ideal gas at density $\rho$ and $\Lambda$ is the De Broglie wavelength.

For any monodisperse hard-sphere packing, whether in equilibrium or not, the void nearest-neighbor functions on the range $[0, \mathcal{D}/2]$ are exactly given by \cite{Re59, To90b}
\begin{subequations}
\begin{equation}
    H_V(r) = \rho s_1(r), \quad r\in [0, \mathcal{D}/2],
    \label{hv_smallr}
\end{equation}
\begin{equation}
    E_V(r) = 1 - \rho v_1(r), \quad r\in [0, \mathcal{D}/2],
\end{equation}
\begin{equation}
    G_V(r) = \frac{1}{1 - \rho v_1(r)}, \quad r\in [0, \mathcal{D}/2].
\end{equation}
\label{hole_stats_smallr}
\end{subequations}
For equilibrium 1D hard rods, the void nearest-neighbor functions for all $r\geq \mathcal{D}/2$ are known exactly \cite{Sa53, To90c}:
\begin{subequations}
\begin{equation}
    H_V(r) = 2\phi \exp\left(\frac{-2\phi (r/\mathcal{D} - 1/2)}{1 - \phi}\right), \quad r \geq \mathcal{D}/2,
\end{equation}
\begin{equation}
    E_V(r) = (1 - \phi) \exp\left(\frac{-2\phi (r/\mathcal{D} - 1/2)}{1 - \phi}\right), \quad r \geq \mathcal{D}/2,
    \label{1d_ev}
\end{equation}
\begin{equation}
    G_V(r) = \frac{1}{1 - \phi}, \quad r \geq \mathcal{D}/2.
\end{equation} 
\label{void_nn_1d}
\end{subequations}
For 2D and 3D equilibrium hard
spheres, it is not possible to obtain exact expressions for the void nearest-neighbor functions for $r\geq \mathcal{D}/2$.
However, accurate approximations for these functions along the disordered branch has been obtained by Torquato \cite{To95a}.
Specifically, for $d = 2$, we have
\begin{equation}
    G_V(r) = a_0 + \frac{a_1}{r}, \quad r \geq \mathcal{D}/2,
    \label{gv2d}
\end{equation}
where the coefficients $a_0$ and $a_1$ are given by \cite{To95a}
\begin{subequations}
\begin{equation}
    a_0 = 
    \begin{cases}
    \frac{1+0.128\phi}{(1 - \phi)^2}, \quad 0 \leq \phi \leq \phi_f,\\
    2g_f(1)\frac{\phi_c - \phi_f}{\phi_c - \phi} - \frac{1}{1 - \phi}, \quad \phi_f \leq \phi \leq \phi_c,
    \end{cases}
\end{equation}
\begin{equation}
    a_1 = 
    \begin{cases}
    \frac{-0.564\phi}{(1 - \phi)^2},  \quad 0 \leq \phi \leq \phi_f,\\
    -g_f(1)\frac{\phi_c - \phi_f}{\phi_c - \phi} + \frac{1}{1 - \phi}, \quad \phi_f \leq \phi \leq \phi_c,
    \end{cases}
\end{equation}
\label{a0a12d}
\end{subequations}
where $\phi_f = 0.69$, $\phi_c = 0.82$, and $g_f(1) = (1 - 0.436\phi_f)/(1 - \phi_f)^2$. 

\section{Biased-sampling Scheme for Hole Statistics}
\label{meth}
Here, we introduce the biased-sampling algorithm that enables one to accurately compute void nearest-neighbor functions for hard-sphere crystals and high-density fluids for radial distances beyond the sphere diameter, i.e., up to rare holes that occur with probabilities at least 5 orders of magnitude lower than the sensitivity limit of the unbiased approach.
Motivated by the work of Zhang and Torquato \cite{Zh17a} to create large holes in particle systems, this algorithm utilizes a test point that interacts with hard spheres via a biasing potential, thereby creating holes that are much larger than those observable in standard simulations.
For clarity, we first describe the algorithm applied to the canonical ensemble, before extending it to the grand canonical ensemble. 
As will be shown in Sec. \ref{res}, finite size effects are significantly reduced in the latter ensemble.

\subsection{Canonical ensemble}
\label{canonical}
We consider a configuration $\mathbf{r}^N$ of $N$ hard spheres of diameter $\mathcal{D}$ in $\mathbb{R}^d$ in a simulation box under periodic boundary conditions of fixed volume $V$. 
To sample the void nearest-neighbor probability distribution function $H_V(r)$, we introduce one additional test point at position vector $\mathbf{t}$, which interacts with its nearest hard-sphere neighbor via a biasing potential $u_i(r)$, where $r = \min_{\mathbf{r} \in \mathbf{r}^N}|\mathbf{r - t}|$ is the distance from $\mathbf{t}$ to its nearest hard-sphere center. 
The potential $u_i(r)$ on the interval $[0, R)$ is subject to iterative optimization to sample hole radii in this range, where $R \leq 2 \mathcal{D}$ is the largest hole radius that we are interested in sampling at a given $\phi$.
In the initial iteration ($i = 0$), we simply use $u_0(r) = 0$, i.e., there is no initial interaction between the test point and the hard spheres.
However, in subsequent iterations ($i \geq 1$), $u_i(r)$ is repulsive on $r \in (\mathcal{D}/2, R)$ due to the updating rule (\ref{uir_update}) described below.
In each iteration $i$,  Monte Carlo (MC) simulations are performed at the dimensionless temperature $k_BT = 1/\beta = 1$ in the canonical ensemble.
Note that both the hard spheres and the test point undergo attempted MC displacements.

Once equilibrium is reached, we use the equilibrated configurations to compute the probability distribution function of the nearest-neighbor distance from the test point to its nearest hard-sphere center, denoted as $p_i(r)$.
To adequately sample this function, a configuration snapshot is taken every 20 sweeps for each simulation trajectory consisting of $2\times 10^5$ sweeps. 
The resulting $p_i(r)$ is obtained from a binned histogram of nearest-neighbor distances to the test point with bin size $\delta = 0.01\mathcal{D}$, averaged over all snapshots in 200 independent simulation trajectories.
The estimate of $H_V(r)$ in iteration $i$ is then given by
\begin{equation}
    H_V^i(r) = \frac{p_i(r)\exp[\beta u_i(r)]}{\int_{\mathbb{R}} p_i(r)\exp[\beta u_i(r)] dr},
    \label{Hvi}
\end{equation}
where the denominator is a normalization factor.

In the next iteration, the biasing potential is updated binwise as
\begin{equation}
    \beta u_{i + 1}(r) = \beta u_i(r) + \ln[p_i(r)\mathcal{D}] + C_i
    \label{uir_update}
\end{equation}
on the range of $r$ where $p_i(r)\mathcal{D} > 10^{-3}$. For $r$ values such that $p_i(r)\mathcal{D} \leq 10^{-3}$, we assume that the statistics for $p_i(r)$ is not sufficiently accurate to properly update $u_i(r)$, and thus a linear extrapolation is used up to $r = 2\mathcal{D}$.
For $r>2\mathcal{D}$, we set $u_i(r) = \infty$, i.e., holes with radii larger than $2\mathcal{D}$ are forbidden.

To facilitate comparison of $u_i(r)$ across iterations, the constants $C_i$ are chosen such that $\beta u_{i}(r)$ at the first bin vanishes for any $i$. 
Note that the values of these constants have no effect on the simulation results.
The iterations are repeated until $p_i(r)$ is nearly uniform within the range of interest $(0, R)$, i.e.,
\begin{subequations}
\begin{equation}
    \frac{1}{R}\int_{0}^{R}(p_i(r)\mathcal{D})^2 d r - \left(\frac{1}{R}\int_{0}^{R}p_i(r)\mathcal{D} d r\right)^2 \leq 0.05, \quad \text{and}
    \label{uniform_condition1}
\end{equation}
\begin{equation}
    p_i(r) > 0.8/R, \quad \forall r\in(0, R).
    \label{uniform_condition2}
\end{equation}
\label{uniform_condition}
\end{subequations}
The convergence criterion (\ref{uniform_condition}) ensures that all nearest-neighbor distances in the range $(0, R)$ are sufficiently sampled.
Specifically, given that $R\leq 2\mathcal{D}$, (\ref{uniform_condition2}) implies that there are at least 8,000 counts in each bin.
We remark that analogs of this condition have been used in various other studies that utilize biased-sampling techniques \cite{Pr01, He04}.

Because the time $\tau$ for the system to reach an equilibrium steady state (in terms of the number of MC sweeps) increases with $\phi$, we choose $R$ such that the hole statistics within $[0, R)$ can be well equilibrated within $2\times 10^6$ sweeps; see Sec. \ref{res:2D} for details.
The system is taken to be well equilibrated if $G_V(r)$ obtained from configurations $10^5$ sweeps apart differ within 3\% at all bins on $[0, R)$.
For 2D and 3D crystals, the magnitude of $-\beta u_i(R)$ in the final iteration is on the order of $10^2$.

\subsection{Grand canonical ensemble}
In contrast to the canonical ensemble, numerical simulation in the grand canonical ensemble facilitates the creation and migration large holes through the addition and removal of particles, thereby significantly reducing finite size effects compared to the canonical ensemble, in which an interstitial has to be created for each vacancy.
In standard grand-canonical simulations without the biasing potential, particle additions at random positions experience a high rejection rate.
By contrast, in the biased-sampling scheme, a large hole centered at $\mathbf{t}$ can be easily created as a result of the biasing potential. 
Therefore, new particles inserted in the vicinity of the test point have a higher probability to be accepted.

In light of this observation, we propose the following method to sample the grand canonical ensemble with chemical potential $\mu$ associated with prescribed mean packing fraction $\phi$.
The initial configurations of our simulations are perfect crystals at $\phi$.
The simulation boxes of volume $V$ under periodic boundary conditions are rhombic in two dimensions and cubic in three dimensions, which are commensurate with 2D triangle-lattice and 3D fcc Bravais lattice vectors, respectively.
In an attempted particle addition, a hard-sphere particle is inserted, the center of which is randomly drawn from the uniform distribution on the spherical region of radius $\lambda - \mathcal{D}$ centered at $\mathbf{t}$, where $\lambda > \mathcal{D}$ is the nearest-neighbor distance to the test point before the addition. 
This is to ensure that the newly added particle becomes the nearest neighbor to the test point and that it does not overlap with any existing hard spheres.
If $\lambda \leq \mathcal{D}$, no particle addition is attempted.
In attempted particle deletions, the nearest hard-sphere neighbor of the test point is removed if its distance to $\mathbf{t}$ is smaller than $\lambda' - \mathcal{D}$, where $\lambda' > \mathcal{D}$ is the second nearest-neighbor distance to the test point prior to the deletion.
If $\lambda' \leq \mathcal{D}$, no particle deletion is attempted.

To maintain detailed balance, the attempted additions and deletions are accepted with probabilities
\begin{subequations}
\begin{equation}
    P_{\text{add}} = \min\left[1, \exp(-\beta \Delta \Phi + \beta\mu')\rho_0 v_1(\lambda - \mathcal{D})\right],
\end{equation}
\begin{equation}
    P_{\text{del}} = \min\left[1, \frac{\exp(-\beta \Delta \Phi - \beta\mu')}{\rho_0 v_1(\lambda' - \mathcal{D})}\right],
\end{equation}
\end{subequations}
where $\Delta\Phi$ is the change in the biasing potential energy caused by the addition or deletion, $\rho_0 = \phi/v_1(\mathcal{D}/2)$ is the mean number density corresponding to the prescribed mean packing fraction and $\mu'$ is the chemical potential in excess of the ideal gas contribution.
In order to relax the full system after an addition or deletion, 50 sweeps of attempted particle displacements are done between each attempted addition or deletion.
Additions and deletions are attempted randomly with equal probabilities.
The iterative computations of $H_V^i(r)$ proceed in the same way as in Sec. \ref{canonical}.

Prior to executing the biased-sampling algorithm in the grand canonical ensemble, one must first find the excess chemical potential $\mu'$.
The chemical potentials for equilibrium hard-sphere systems in one, two and three dimensions have been reported in various previous works across packing fractions \cite{St65, To90b, Ro12, Us17}.
For 1D hard rods, $\mu'$ is exactly known from Eqs. (\ref{muprime_ev}) and (\ref{1d_ev}), i.e.,
\begin{equation}
    \beta \mu'_{d = 1}(\phi) = -\ln[E_V(\mathcal{D})] = - \ln(1 - \phi) + \frac{\phi}{1 - \phi}.
\end{equation}
For 2D hard-disk and 3D hard-sphere crystals, we use the approximation obtained by Stillinger et al. \cite{St65}:
\begin{equation}
    \beta \mu'_d(\phi) = -d\ln(1 - \frac{\phi}{\phi_J}) + \frac{\beta p(\phi; d)}{\rho} = -d\ln(1 - \frac{\phi}{\phi_J}) + 
    \frac{d}{1 - \phi/\phi_J(d)},
    \label{mu_2D3D}
\end{equation}
where $\phi_J(2) = \pi/\sqrt{12}$ and $\phi_J(3) = \pi/\sqrt{18}$ are the close-packing fractions of the 2D triangle lattice and the 3D fcc crystals, respectively, and $\beta p(\phi, d)/\rho = d/(1 - \phi/\phi_J(d))$ is the reduced pressure approximated via free-volume theory \cite{Sa62}.
Note that Eq. (\ref{mu_2D3D}) agrees very well with numerical results for $\mu'$ via density functional theory \cite{Ro12} and kinetic Monte Carlo \cite{Us17} calculations.

It is not known if Eq. (\ref{mu_2D3D}) provides an accurate approximation for the excess chemical potential in the 2D hexatic phase. 
Thus, in this case, we numerically determine $\mu'$ following the gradual insertion method developed by Mon and Griffiths \cite{Mo85}, which is an extension of the well-established Widom insertion method \cite{Wi82} to treat high-density systems in which direct particle insertions experience a high rejection rate. 
In this method, a pseudo hard sphere at position vector $\mathbf{p}$ is inserted into a configuration $\mathbf{r}^N$ of $N$ hard spheres, and it interacts with the existing hard-sphere \textit{centers} via a pseudo hard-sphere potential $v(r; \sigma)$, where $\sigma$ is the effective hard-sphere diameter.
This system is equilibrated using Monte-Carlo simulations, while $\sigma$ is increased gradually from $0$ to $\mathcal{D}$ in the canonical ensemble.
The excess chemical potential is given by 
\begin{equation}
    \mu' = \int_0^\mathcal{D} \langle\frac{d\Psi(\sigma)}{d\sigma}\rangle_{N + \sigma} d\sigma,
    \label{mu_integral}
\end{equation}
where $\langle x \rangle_{N + \sigma}$ is the mean of quantity $x$ in the canonical ensemble of $N$ hard spheres and the additional pseudo hard sphere with pair interaction $v(r; \sigma)$, and
\begin{equation}
\Psi(\sigma) = \Sigma_{\mathbf{r}_i \in \mathbf{r}^N} v(|\mathbf{r}_i - \mathbf{p}|; \sigma)
\end{equation}
is the potential energy experienced by the pseudo hard sphere.
In this work, we use the following pseudo hard-sphere potential
\begin{equation}
    \beta v(r; \sigma) = \left(\frac{4}{(r/\sigma)^{50}}\right)\left(\frac{-\operatorname{atan}(\frac{r/\sigma - 1}{0.001})}{\pi} + \frac{1}{2}\right).
    \label{pseudoHS}
\end{equation}
Equation (\ref{pseudoHS}) has the same repulsive part (up to scaling) as the well-known generalized Lennard-Jones potential \cite{Jo12}, but has the additional advantage that $v(r; \sigma)$ decreases to zero rapidly due to the presence of the arctangent function.
The integral (\ref{mu_integral}) is evaluated via the trapezoidal rule using 20 equal-sized bins, while the quantity $d\Psi(\sigma)/d\sigma$ is computed via automatic differentiation \cite{Re16b}. 

\section{Results for the Hole Statistics of Equilibrium Hard Spheres}
Here, we present the results for the void nearest-neighbor functions obtained using our biased-sampling scheme described in Sec. \ref{meth}. 
Since $H_V(r), E_V(r)$ and $G_V(r)$ can be easily converted to one another via Eqs. (\ref{def_ev}) and (\ref{def_gv}), we report here only results for $G_V(r)$ due to its direct relationship to the reduced pressure (\ref{pressure}) and the fact that plots of $G_V(r)$ are more visually intuitive than those of $H_V(r)$ and $E_V(r)$ \cite{Wa20}.
\label{res}
\subsection{1D Hard-Rod Fluids}
\begin{figure*}[]
\subfloat[]{\includegraphics[width = 60mm]{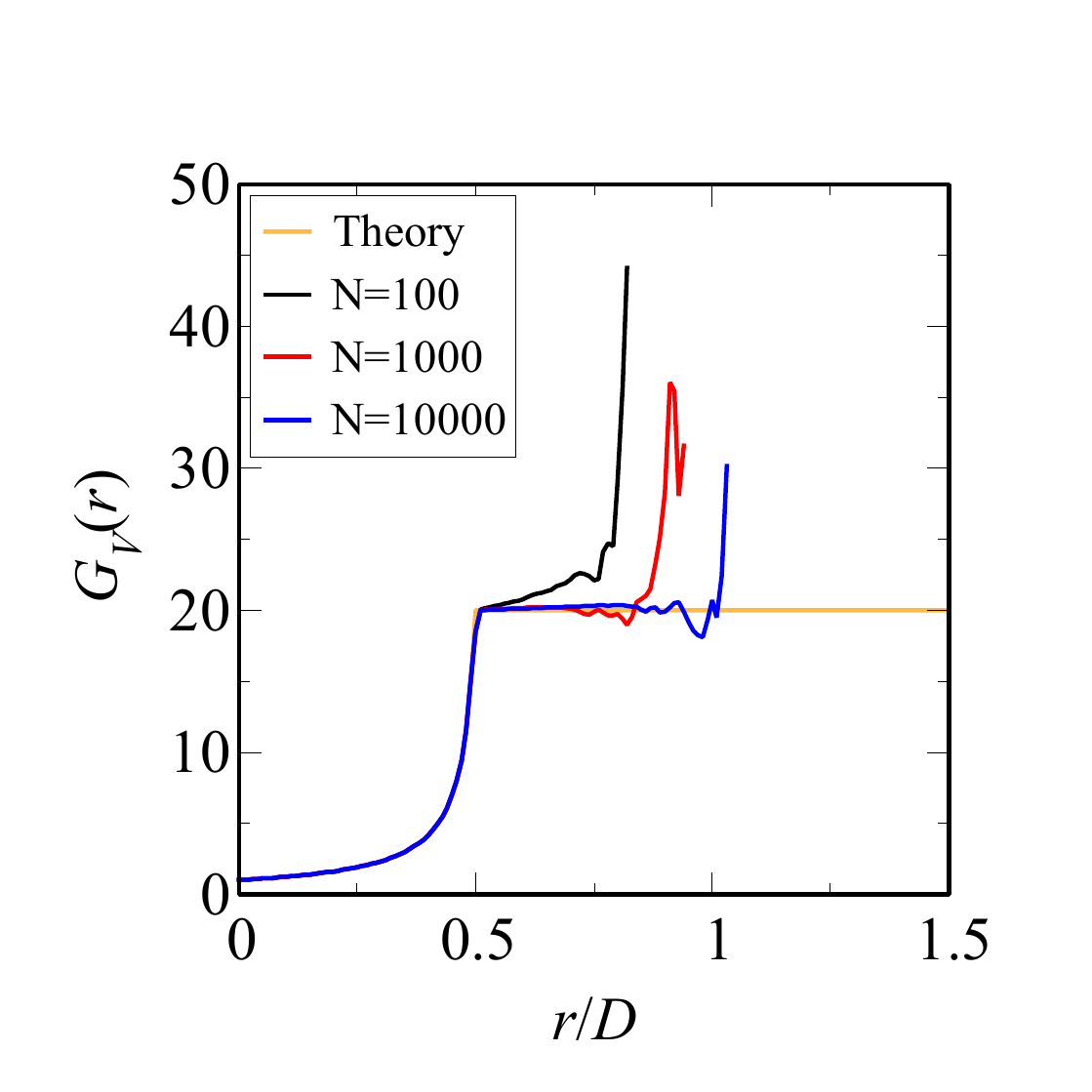}}
    \subfloat[]{\includegraphics[width = 60mm]{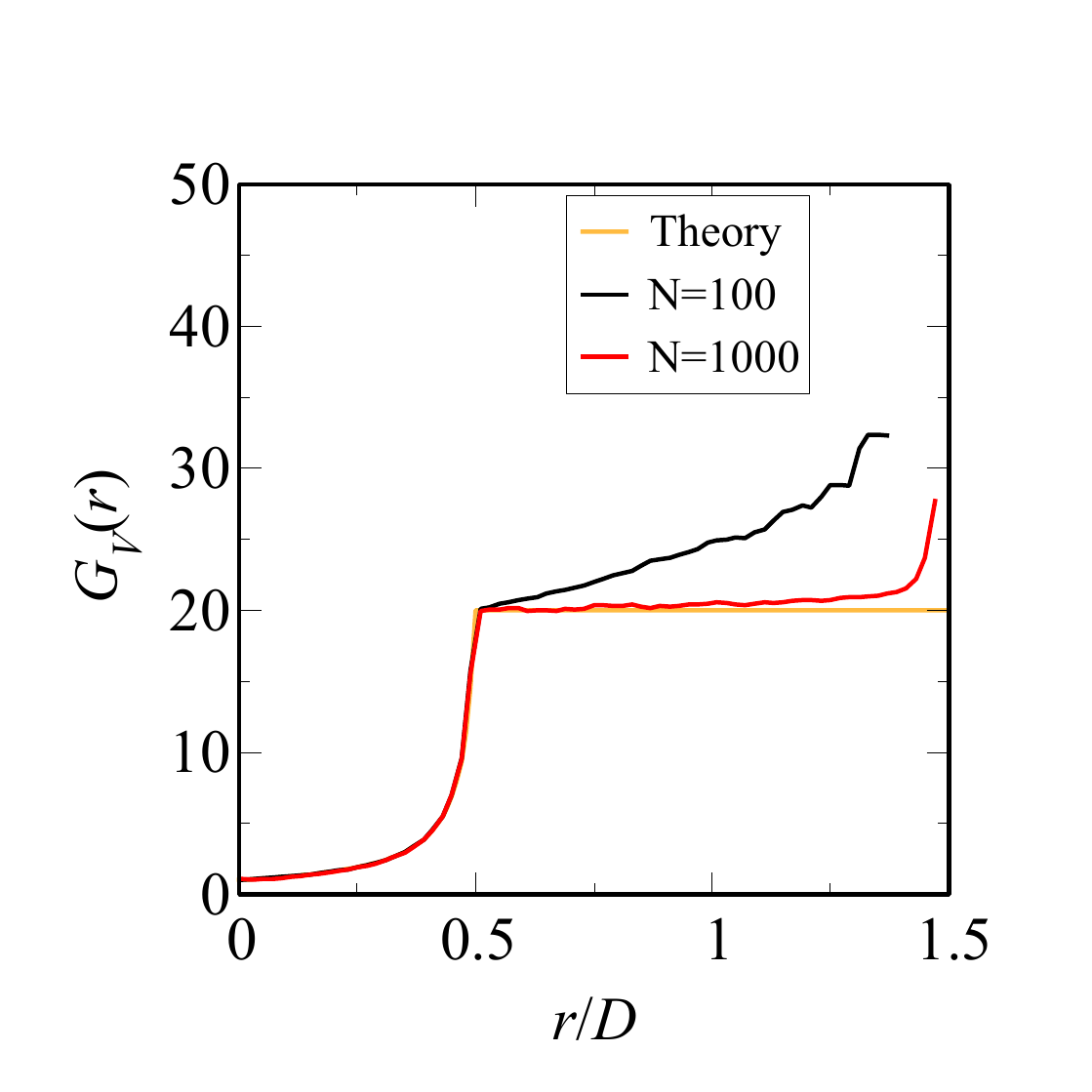}}
    \subfloat[]{\includegraphics[width = 60mm]{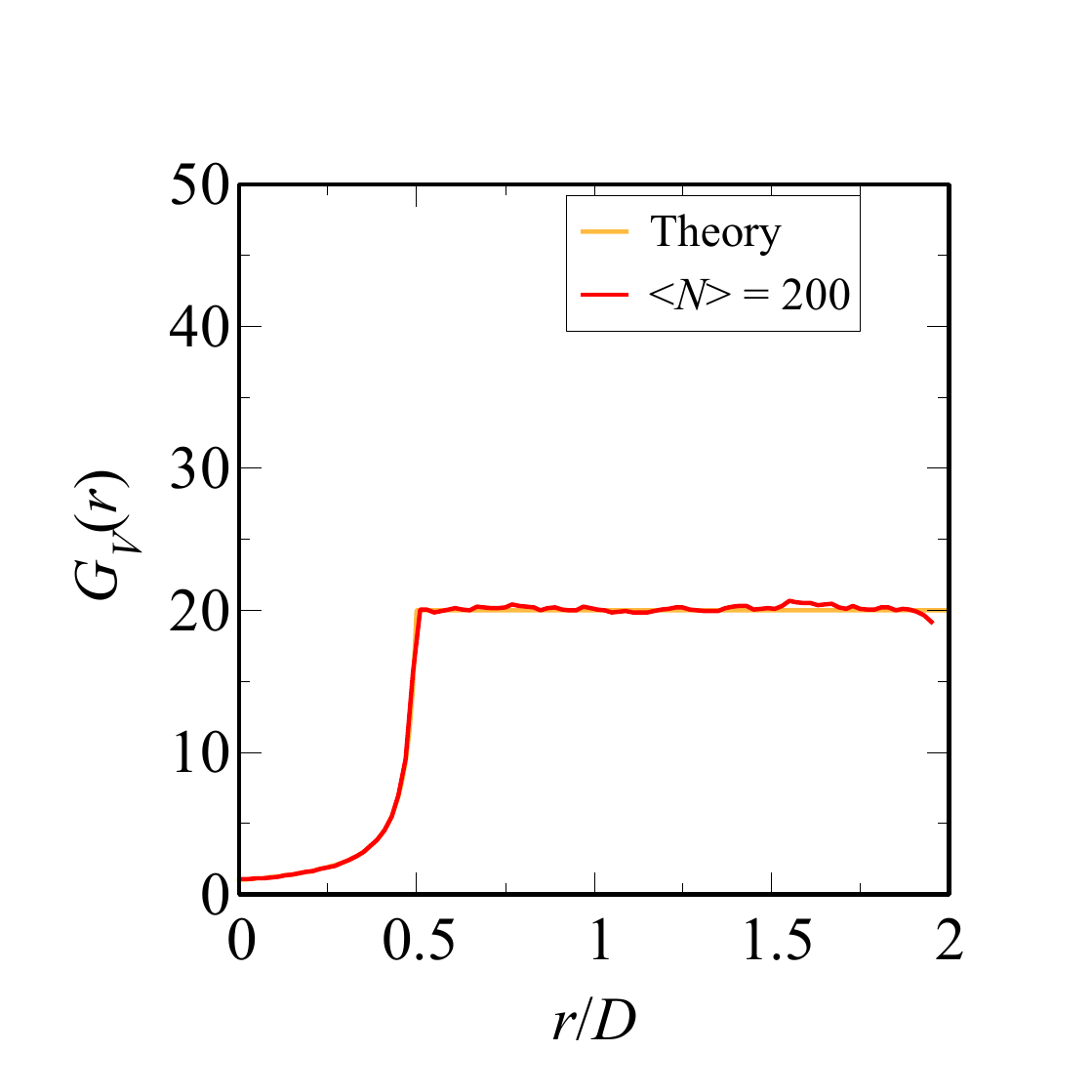}}
    \caption{Conditional void probability function $G_V(r)$ for equilibrium 1D hard-rod fluid at $\phi = 0.95$ computed via various methods. (a) $G_V(r)$ computed via the standard sampling method in the canonical ensemble. (b) $G_V(r)$ computed via biased sampling in the canonical ensemble. (c) $G_V(r)$ computed with biased sampling in the grand canonical ensemble.}
    \label{fig:1DHSphi0x95}
\end{figure*}

To demonstrate the validity and accuracy of our biased-sampling scheme, we apply it to compute the void nearest-neighbor functions for equilibrium 1D hard-rod fluids, whose functional forms in the thermodynamic limit are exactly known, as given by Eqs. (\ref{hole_stats_smallr}) and (\ref{void_nn_1d}). 
Figure \ref{fig:1DHSphi0x95} shows results for $G_V(r)$ for the 1D hard-rod fluid at prescribed packing fraction $\phi = 0.95$.
As shown in Fig. \ref{fig:1DHSphi0x95}(a), the standard sampling approach in the canonical ensemble requires a large number of particles ($N = 10,000$) to achieve reasonable accuracy (3\% error) on the range $[0, \mathcal{D}]$. 
Note that all of the simulated $G_V(r)$ increase sharply at a certain hole radius $r^*$, which indicates that in standard sampling, one can not detect holes with radii larger than $r^*$.
By contrast, Fig. \ref{fig:1DHSphi0x95}(b) shows that with biased-sampling in the canonical ensemble, one can accurately determine hole statistics for larger hole radii, up to $r = 1.5\mathcal{D}$, using much smaller system sizes compared to standard simulations. 
Due to the finite-size effect caused by fixed $N$ and $V$ in the canonical ensemble, $G_V(r)$ obtained via such simulations increase with $r$ on the range $r\geq \mathcal{D}/2$.
However, as $N$ increases, the corresponding $G_V(r)$ approaches the expected value $20 = 1/(1 - \phi)$ in the thermodynamic limit.
Figure \ref{fig:1DHSphi0x95}(c) shows that the biased-sampling scheme in the grand canonical ensemble is able to very accurately determine $G_V(r)$ up to $r = 1.95\mathcal{D}$ using merely $\langle N\rangle=200$ particles. 
Compared to biased sampling in the canonical ensemble, the grand canonical ensemble further improves the sensitivity and accuracy of hole sampling, because holes can be readily created via the deletion of particles.
Importantly, due to the strong finite-size effect associated with the canonical ensemble, we subsequently exclusively use the grand canonical ensemble in our large-scale calculations for 2D and 3D crystal states.

\subsection{Computational Complexity for Crystal States}
\label{sec:complexity}
\begin{figure}[htp]
    \centering
    \subfloat[]{\includegraphics[width=80mm]{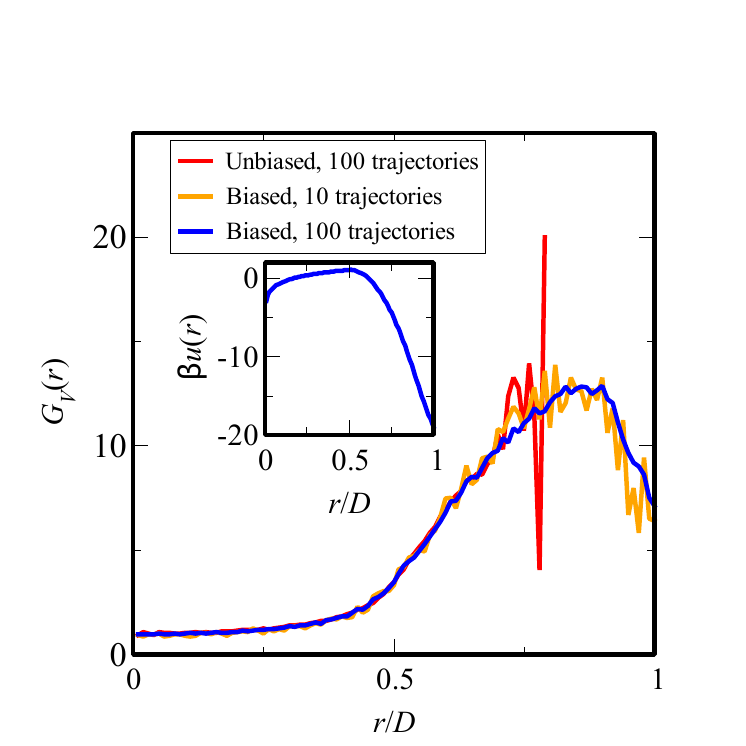}}
    \subfloat[]{\includegraphics[width=80mm]{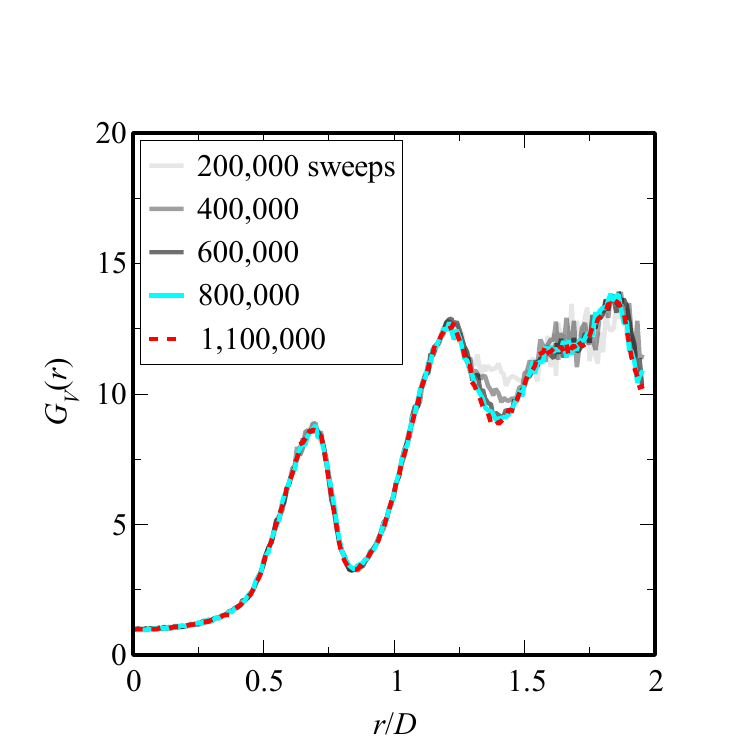}}
    
    \subfloat[]{\includegraphics[width=77.5mm, trim = {0, -0.45cm, 0, 0}, clip]{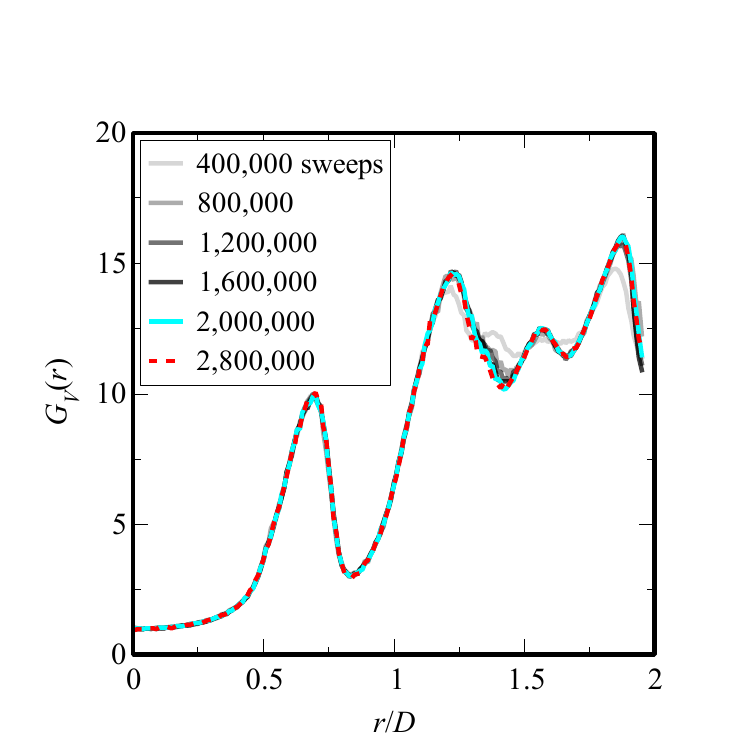}}
    \subfloat[]{\includegraphics[width=80mm]{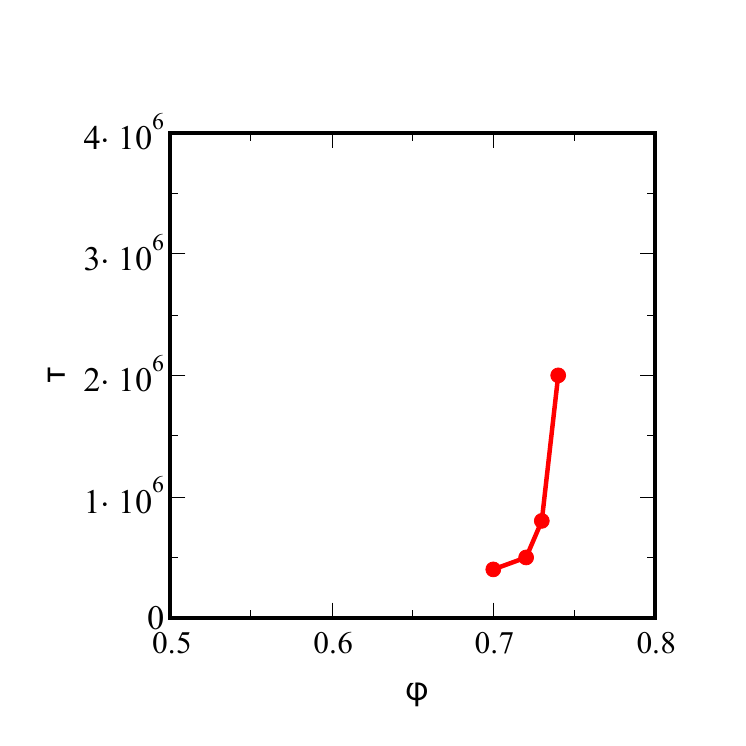}}
    \caption{
    (a) Comparison of $G_V(r)$ computed via the unbiased and biased-sampling approaches for the 2D equilibrium hard-disk crystal at $\phi = 0.73$ in the \textit{canonical} ensemble.
    The inset shows the corresponding biasing potential.
    (b)--(c) The evolution of $G_V(r)$ for the 2D equilibrium hard-disk crystal at prescribed mean packing fractions (b) $\phi = 0.73$ and (c) $\phi = 0.74$ under the biased-sampling scheme in the \textit{grand canonical} ensemble. 
    (d) The equilibration time $\tau$ for $G_V(r)$ on the range $[0, 1.5\mathcal{D})$ at $\phi = 0.7, 0.72, 0.73$, and $0.74$ in the grand canonical ensemble.}
    \label{fig:complexity}
\end{figure}

To ascertain the accuracy and efficiency of our biased-sampling method to study the hole statistics of 2D and 3D crystal states, we first perform a preliminary study by computing $G_V(r)$ for the 2D equilibrium hard-disk crystal at $\phi = 0.73$ in the \textit{canonical} ensemble using both the biased and the standard (unbiased) methods, as shown in Fig. \ref{fig:complexity}(a). 
Here, the biasing potential (plotted in the inset) is determined via the iterative procedure in Sec. \ref{meth} such that $p_i(r)$ is approximately uniform on the range $[0, \mathcal{D}]$ [see Eq. (\ref{uniform_condition})].
Each simulation trajectory, in which 2500 configurations with $N = 256$ particles are generated, takes around 30 min to complete on a 2.8 GHz Intel CPU.
We find that with 100 trajectories, $G_V(r)$ via the standard unbiased method has significant noise at $r \sim 0.75\mathcal{D}$, and no hole with radii larger than $0.80\mathcal{D}$ is found.
This limitation is due to the fact that $E_V(0.8\mathcal{D}) \sim 10^{-5}$, which is the smallest probability that standard simulations on this scale can detect.
By contrast, the biased-sampling method reaches comparative precision with the unbiased method at $r \sim 0.75\mathcal{D}$ with as few as 10 trajectories and can detect larger holes with radii up to $\mathcal{D}$.
For 100 trajectories, the unbiased method produces accurate hole statistics on the range of interest $[0, \mathcal{D}]$.
Notably, according to our biased-sampling result, $E_V(\mathcal{D})\sim 10^{-10}$, which is $10^{-5}$ of the sensitivity limit of the unbiased method.
Thus, it would require the unbiased method on the order of $10^5$ hours to reach comparative accuracy at $r \sim \mathcal{D}$ with the biased-sampling method.
We remark that in this preliminary investigation, the canonical ensemble is used merely to facilitate the comparison between biased and unbiased algorithms. 
In our subsequent large-scale calculations (200 trajectories of $10^4$ configurations), we exclusively use biased-sampling in the grand canonical ensemble.

Prior to performing large-scale simulations for 2D and 3D crystal states, we first determine the ranges $[0, R)$ on which one can accurately sample the hole statistics at given values of $\phi$ within $2\times 10^6$ sweeps, which takes approximately 24 hours for $\sim250$ particles in two dimensions.
To determine $R$, we estimate the equilibration time $\tau$ required to achieve the convergence condition in Sec. \ref{meth} using the biased-sampling scheme in the grand canonical ensemble for 2D crystals slightly above the hexatic-solid transition packing fraction $\phi_h = 0.72$ \cite{Ber11}.
Figure \ref{fig:complexity}(b) and (c) show the evolution of $G_V(r)$ for the 2D crystal at $\phi = 0.73$ and 0.74, respectively, under their corresponding final biasing potentials, starting from the perfect triangle-lattice configurations at the prescribed mean packing fractions in rhombic simulation cells with $N = 256$.
At $\phi = 0.73$, $G_V(r)$ on the range $[0, 1.25\mathcal{D})$ reaches a steady state within the first 200,000 sweeps.
The range $[1.25\mathcal{D}, 1.95\mathcal{D})$ is equilibrated within the first 800,000 sweeps. 
The significant increase of the equilibration time as $r$ increases is due to the low transition rate between the first, second and third stable hole sizes, with $r \sim \mathcal{D}, 1.5\mathcal{D}$ and $2\mathcal{D}$, respectively.
While the creation of large holes is favored energetically due to the biasing potential, the dynamics of such processes is slow, as particles in the vicinity of the test point must undergo large displacements compared to the mean displacement magnitudes in the unbiased equilibrium crystal.
At $\phi = 0.74$, we find that $2\times 10^6$ sweeps are required to equilibrate $G_V(r)$ up to $1.5\mathcal{D}$.
Due to the dramatic increase of equilibration time as $\phi$ increases, as shown in Fig. \ref{fig:complexity}(d), we choose decreasing values of $R$ as $\phi$ increases, and our choices of $R$ at each 2D and 3D packing fraction are listed in Table \ref{tab:bigR}.
Except in the case $d = 2, \phi = 0.73$, we always discard the data between $1.95\mathcal{D}$ and $2\mathcal{D}$, because the imposed infinite biasing potential at $r = 2\mathcal{D}$ means that no holes with radii large than $2\mathcal{D}$ can be created, which causes an anomalous sharp increase in $G_V(r)$ just before $2\mathcal{D}$.
However, in just one case $d = 2, \phi = 0.73$, we set the hard wall in the biasing potential to be at $2.1 \mathcal{D}$ rather than $2\mathcal{D}$, so that the positions of the local extrema in $G_V(r)$ up to $2\mathcal{D}$ can be accurately determined.
We select this packing fraction since it lies fully in the crystal phase and its corresponding $G_V(r)$ can be readily computed in $2\times 10^6$ sweeps.
We find that in this case, errors in $G_V(r)$ on $[0, 2\mathcal{R}]$ are within 5\%. 
Notably, $G_V(r)$ on the range $[0, 1.1\mathcal{D})$ is equilibrated within the first 200,000 sweeps for all 2D cases with $\phi \leq 0.85$ and 3D cases with $\phi \leq 0.60$.
The total computation time for all biased-sampling calculations is approximately 300 h under the respective final biasing potentials with 200 parallel threads.
In addition, the computation resources needed to evolve the biasing potential prior to convergence is 3--10 times of the final simulation time.
We remark that at any $\phi$ value listed in Table \ref{tab:bigR}, the biased-sampling scheme improves the sensitivity limit, and thus the computational time for reaching comparable accuracy, by at least 5 orders of magnitude compared to the unbiased method.

\begin{table}[htp]
    \centering
    \caption{The upper radius below which $G_V(r)$ can be determined within $\pm 3\%$ error in $2\times 10^6$ sweeps.}
    \begin{tabular}{||c|c||c|c||}
       $\phi$ for 2D states  & $R/\mathcal{D}$ & $\phi$ for 3D states & $R/\mathcal{D}$ \\
       \hline
       $\leq 0.72$ & 1.95 & 0.54 & 1.5 \\
       0.73  & 1.95; Errors are within $5\%$ with $R = 2\mathcal{D}$. & $0.55, 0.60$ & 1.1 \\
       0.74 & 1.5 & & \\
       0.75 & 1.3 & & \\
       $0.76, 0.78, 0.85$ & 1.1 & & \\
    \end{tabular}
    \label{tab:bigR}
\end{table}

\subsection{2D Hard-Disk States}
\label{res:2D}
Here, we present the results for the void nearest-neighbor functions for 2D equilibrium hard-disk crystals, as well as those for a high-density liquid and a hexatic state. 
Furthermore, we study the correlation between large-$r$ behaviors of $G_V(r)$ and the local coordination geometry of holes, which could not be done in previous works due to the absence of accurate hole statistics for $r\geq \mathcal{D}$. 

\subsubsection{Crystal states}

\begin{figure}[htp]
    \centering
    \subfloat[]{\includegraphics[width=80mm]{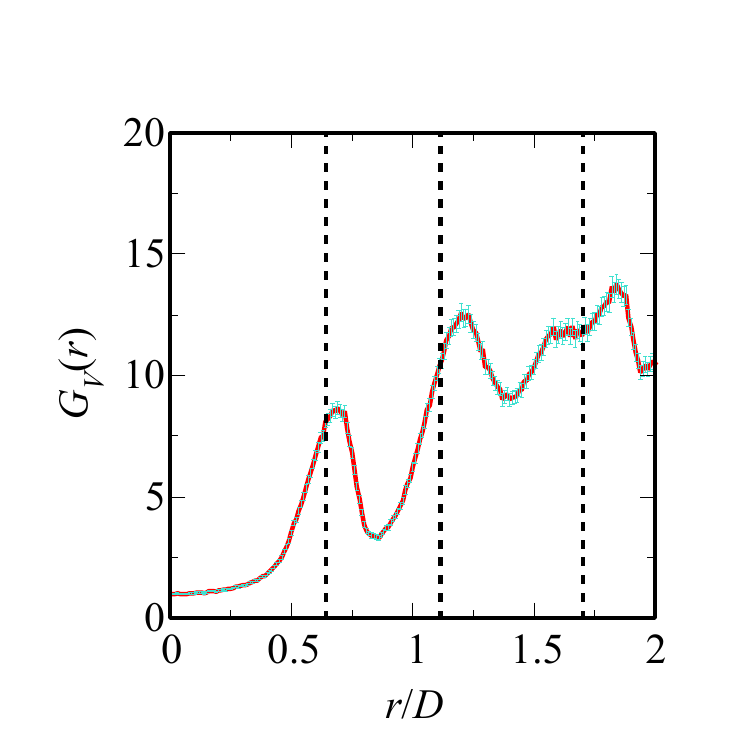}}
    \subfloat[]{\includegraphics[width=80mm]{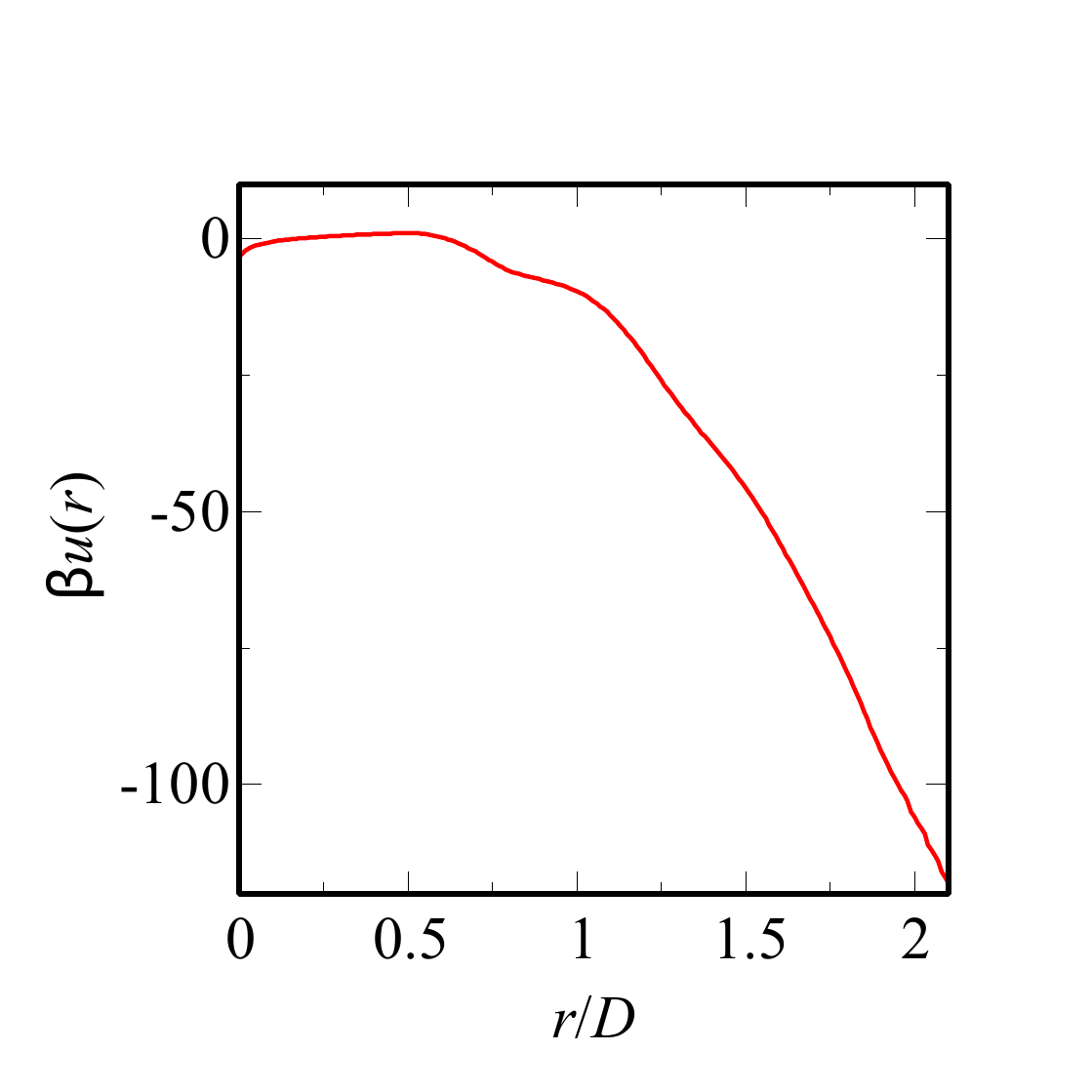}}
    \caption{(a) The function $G_V(r)$ for the 2D equilibrium hard-disk crystal at $\phi = 0.73, \phi/\phi_J = 0.805$ obtained from the grand-canonical biased-sampling scheme. 
    The vertical lines at $r = 0.6435\mathcal{D}, 1.115\mathcal{D}$ and $1.702\mathcal{D}$ correspond to the radii of the circumscribed circles of the idealized holes in Fig. \ref{fig:hole_idealized}(a)--(c), which provide lower bounds of the first three maxima in $G_V(r)$.
    (b) The final biasing potential $u(r)$ used in the biased-sampling scheme in the grand canonical ensemble to compute the void nearest-neighbor functions for the 2D equilibrium hard-disk crystal at $\phi = 0.73$.}
    \label{fig:2DHSphi0x73}
\end{figure}

Figure \ref{fig:2DHSphi0x73}(a) shows simulated results of $G_V(r)$ for the 2D hard-disk crystal at prescribed mean packing fraction $\phi = 0.73$, which is slightly above the packing fraction $\phi_h = 0.72$ corresponding to the second-order phase transition between the hexatic phase and the crystal phase \cite{Ber11}. 
The grand-canonical biased-sampling scheme enables us to determine accurate $G_V(r)$ within $\pm 5\%$ error up to $r = 2\mathcal{D}$.
We note that $E_V(2\mathcal{D})\sim 10^{-48}$, which is more than 40 orders of magnitude smaller than the sensitivity limit of the unbiased method described in Sec. \ref{intro}.
Importantly, we find, for the first time, that the simulated $G_V(r)$ for the crystal state exhibits significant oscillations. 
Such oscillations persist on a large range of $r$ up to $r = 2\mathcal{D}$, and the period of the oscillations is approximately $0.5\mathcal{D}$.
Figure \ref{fig:2DHSphi0x73}(b) plots the final biasing potential $u(r)$ used in the grand-canonical biased-sampling scheme to compute the void nearest-neighbor functions for the 2D equilibrium hard-disk crystal at $\phi = 0.73$.
Note that $u(r)$ is an increasing function with $r$ on $[0, \mathcal{D}/2]$, because $H_V(r)$ increases on this range; see (\ref{hv_smallr}). Thus, $u(r)$ must increase with $r$ to sample these small holes uniformly. 
However, $u(r)$ for $r$ larger than the disk radius is strongly repulsive, so that large holes up to $r = 2\mathcal{D}$ can be adequately sampled.
We remark that, up to an additive constant, the functional form of $u(r)$ is very close to that of $\ln[H_V(r)]$ on the range of interest $[0, 2\mathcal{D}]$, as expected from the updating rule (\ref{uir_update}) and the convergence criterion (\ref{uniform_condition}).
Indeed, the difference between $u(r)$ and $\ln[H_V(r)]$ is within 2\% for all $r$ on the range of interest.
Thus, the oscillations in $u(r)$ imply that $\ln[H_V(r)]$ also contains such oscillations.

\begin{figure}[htp]
    \subfloat[]{\includegraphics[width = 50mm]{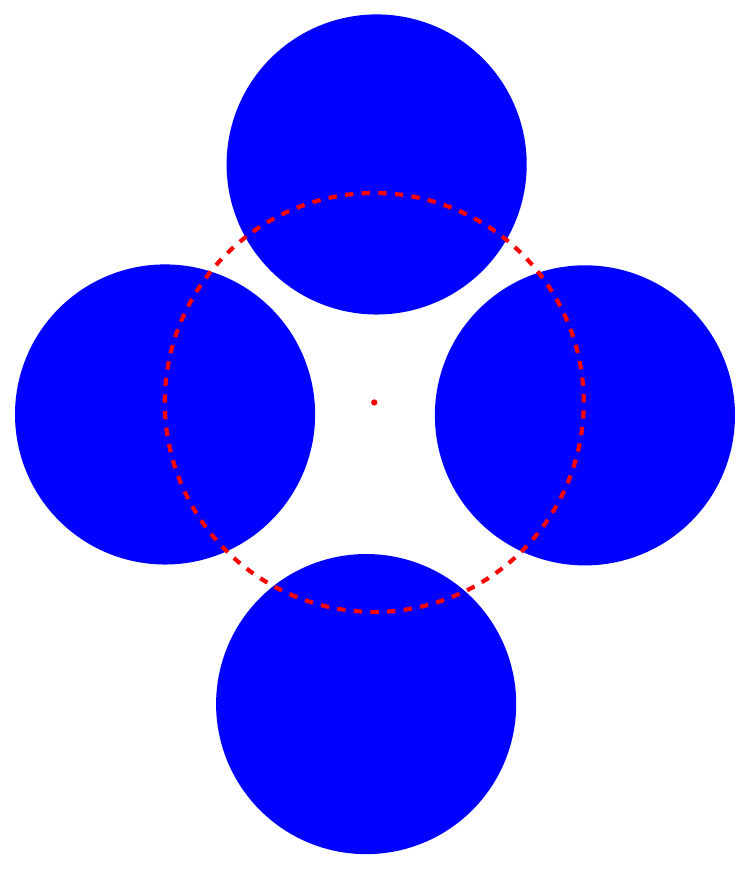}}
    \subfloat[]{\includegraphics[width = 50mm]{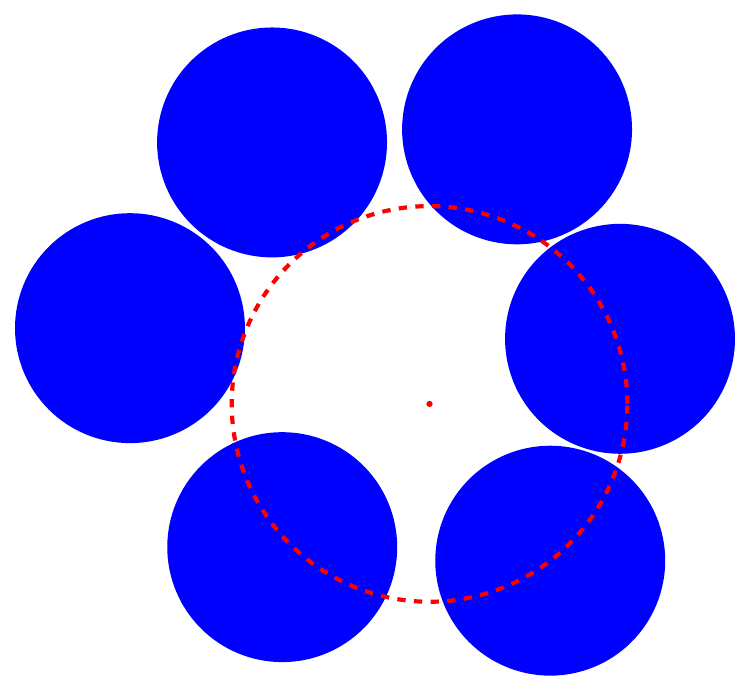}}
    \subfloat[]{\includegraphics[width = 50mm]{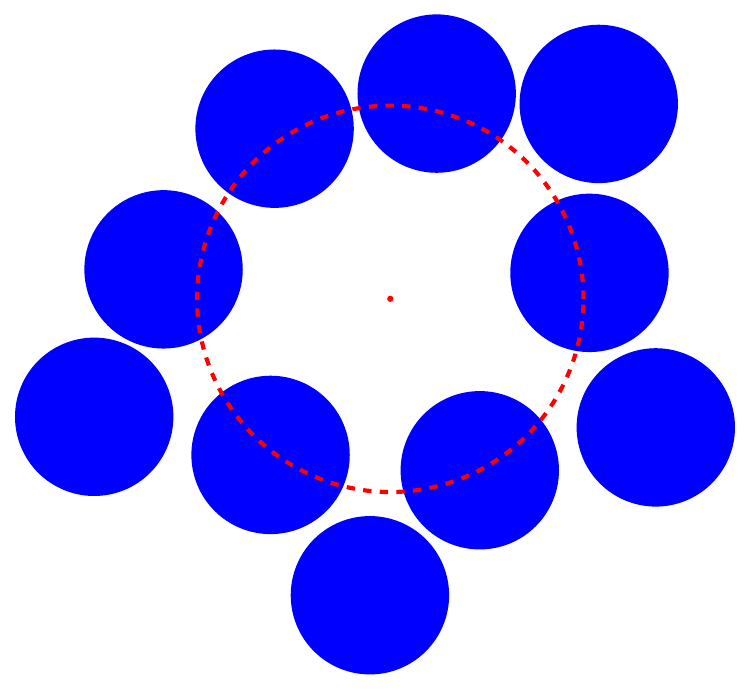}}
    
    \subfloat[]{\includegraphics[width = 50mm]{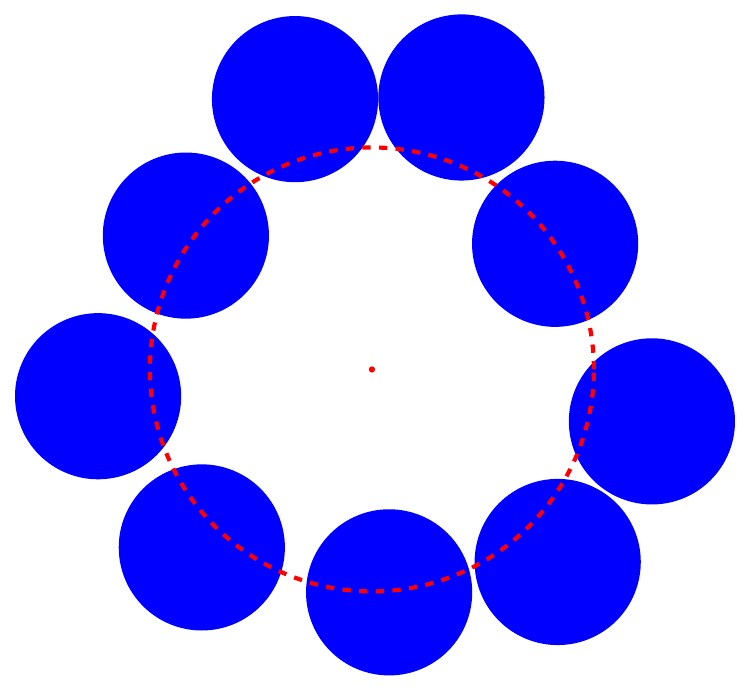}}
    \subfloat[]{\includegraphics[width = 50mm]{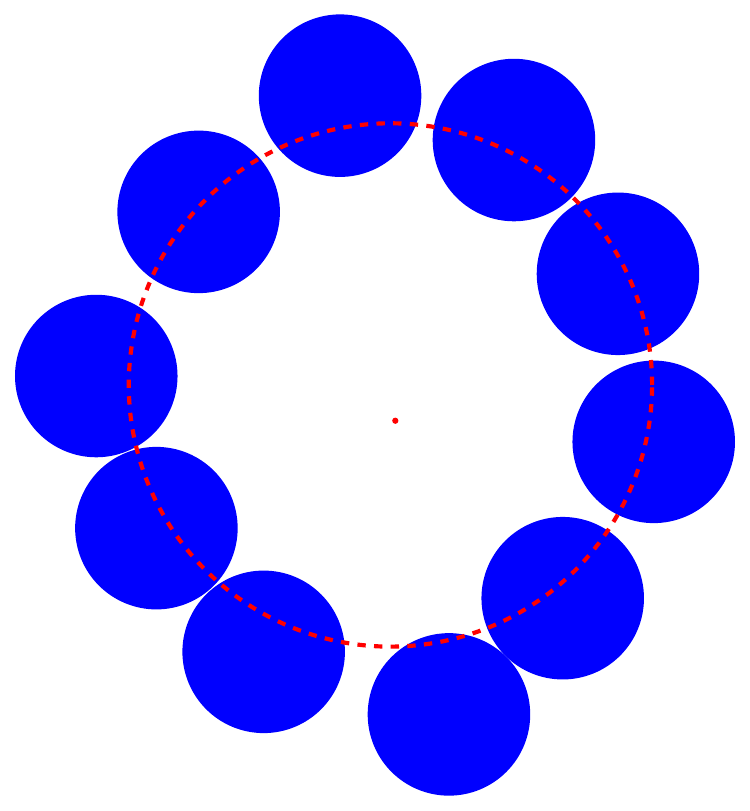}}
    \subfloat[]{\includegraphics[width = 50mm]{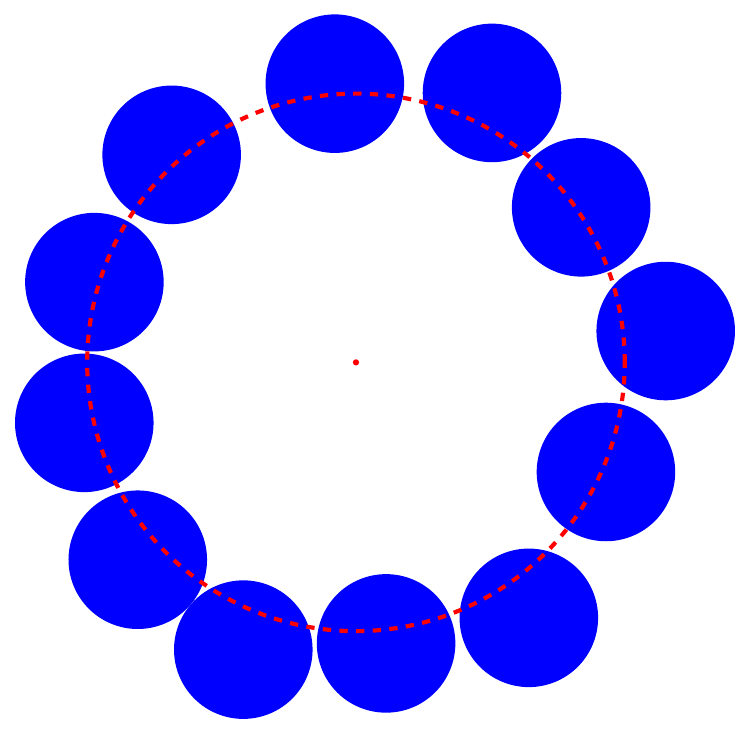}}
    \caption{
    Local coordination configurations of circular-like holes in an equilibrium hard-disk crystal at $\phi = 0.73$. 
    (a) A hole of radius $r=0.69\mathcal{D}$, corresponding to a local maximum in $G_V(r)$.
    (b) A hole of radius $r=0.86\mathcal{D}$, corresponding to a local minimum in $G_V(r)$.
    (c) A hole of radius $r=1.20\mathcal{D}$, corresponding to a local maximum in $G_V(r)$.
    (d) A hole of radius $r=1.37\mathcal{D}$, corresponding to a local minimum in $G_V(r)$.
    (e) A hole of radius $r=1.61\mathcal{D}$, corresponding to a shoulder in $G_V(r)$.
    (f) A hole of radius $r=1.95\mathcal{D}$, corresponding to a local minimum in $G_V(r)$.
    }
    \label{fig:hole_geom}
\end{figure}

To investigate the origin of the oscillations in $G_V(r)$ for crystals, we study the local coordination geometry of circular-like holes in the 2D crystal configurations.
We exclusively consider 2D circular-like holes, because $G_V(r)$ explicitly samples $d$-dimensional spherical holes. 
Indeed, we find that $G_V(r)$ is relatively insensitive to aspherical holes.
Here, we define the coordination number $Z$ of a hole of radius $r$ as the number of disks that overlap with the circle of radius $r$ centered at the test point.
Figure \ref{fig:hole_geom} shows representative coordination geometries in the equilibrium hard-disk crystal at $\phi = 0.73$.
The first local maximum in $G_V(r)$, occurring at $r = 0.69\mathcal{D}$, corresponds to 4-coordinated holes [Fig. \ref{fig:hole_geom}(a)].
As expected, this hole radius is approximately half of the diagonal length of a square with side length $\mathcal{D}$.
The first local minimum in $G_V(r)$ at $r = 0.86\mathcal{D}$ corresponds to hexagonal holes with $Z = 6$ [Fig. \ref{fig:hole_geom}(b)].
Figure \ref{fig:hole_geom}(c) shows that the second local maximum in $G_V(r)$, occurring at $r = 1.20 \mathcal{D}$, corresponds to a significantly distorted hexagonal hole, i.e., the hexagon formed by the centers of the nearest 6 hard disks to the test point is distinctly different from a regular hexagon.
Figures \ref{fig:hole_geom}(d) and (f) show that the second and third local minima in $G_V(r)$, at $r = 1.37 \mathcal{D}$ and $r = 1.95 \mathcal{D}$, respectively, correspond to 9- and 12-coordinated holes that are created by removing 3 and 7 particles in the triangle lattices, respectively.
We also find that the shoulder of $G_V(r)$ at $r\sim 1.65\mathcal{D}$ corresponds to 10-coordinated holes [Fig. \ref{fig:hole_geom}(e)].
These observations suggest that the minima in $G_V(r)$ correspond to  ``stable'' hole radii are compatible with the local configuration of the triangle-lattice crystal, whereas maxima in $G_V(r)$ correspond to hole sizes that lead to significant distortion of the triangle lattice and represent the transition state between stable hole sizes. 

\begin{figure}[htp]
    \subfloat[]{\includegraphics[width = 60mm]{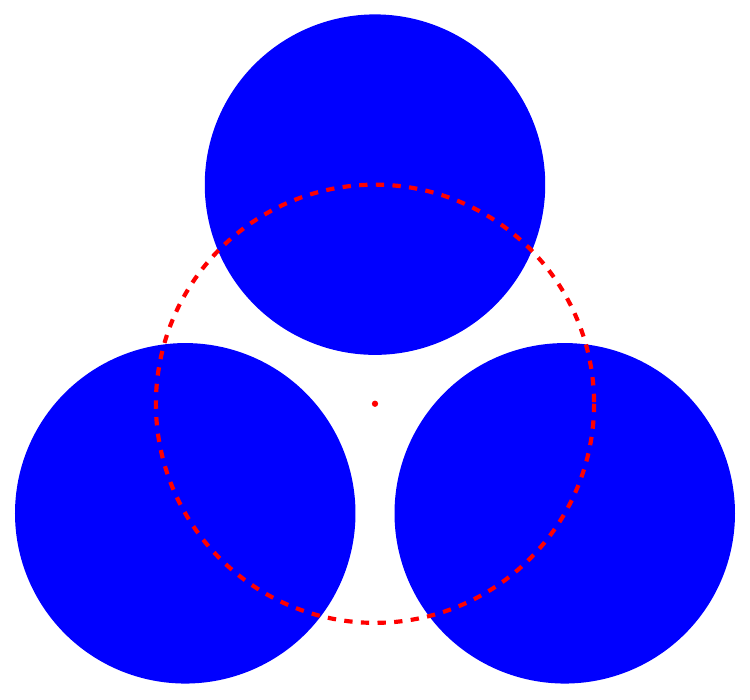}}
    \subfloat[]{\includegraphics[width = 60mm]{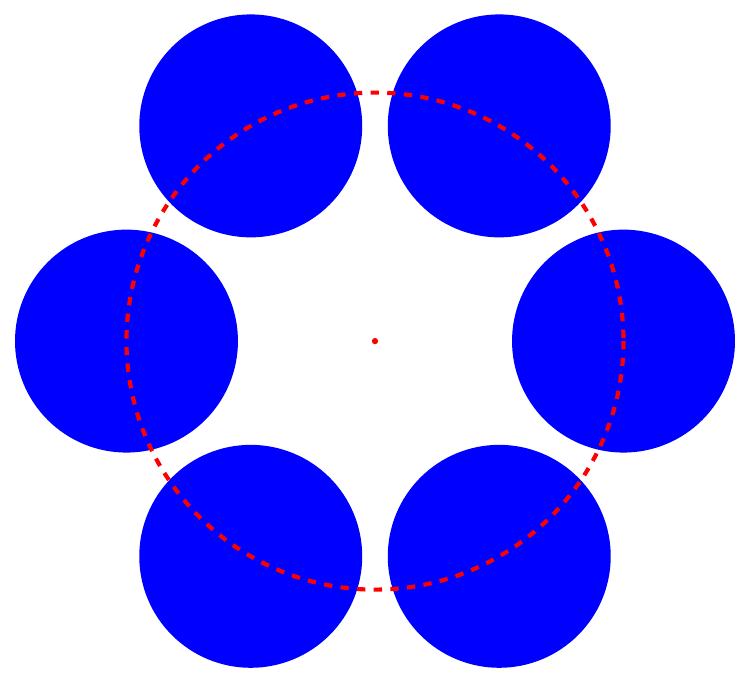}}
    
    \subfloat[]{\includegraphics[width = 60mm]{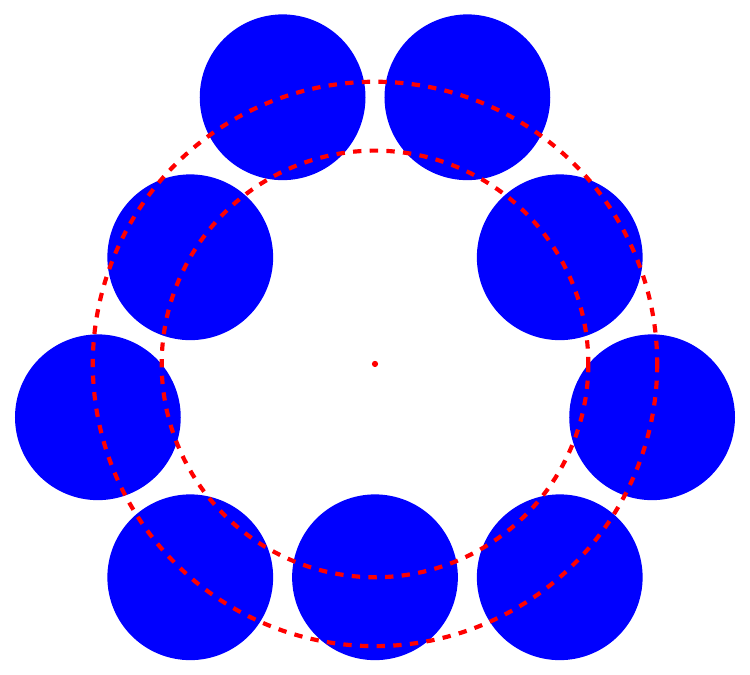}}
    \subfloat[]{\includegraphics[width = 60mm]{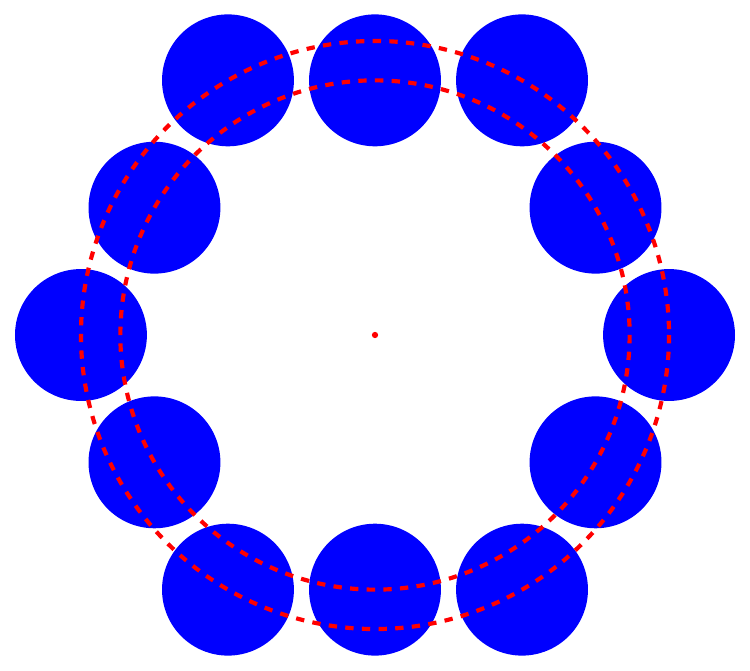}}
    \caption{
    Local coordination configurations of circular-like holes in a perfect hard-disk crystal at $\phi = 0.73$. 
    (a) A triangular hole of radius $r=0.6435\mathcal{D}$, which is the covering radius of the perfect crystal.
    (b) A hexagonal hole of radius $r=1.115\mathcal{D}$.
    (c) A 9-coordinated hole with radius of the inscribed circle $r=1.287\mathcal{D}$ and radius of the circumscribed circle $r=1.702\mathcal{D}$.
    (d) A 12-coordinated hole of radius of the inscribed circle $r=1.931\mathcal{D}$ and radius of the circumscribed circle $r=2.229\mathcal{D}$.
    }
    \label{fig:hole_idealized}
\end{figure}

To explain quantitatively the positions of the maxima in $G_V(r)$, we compare the local coordination geometry of the stable holes in equilibrium hard-disk systems at $\phi = 0.73$ with the corresponding idealized holes in a perfect triangle lattice, as shown in Fig. \ref{fig:hole_idealized}.
The idealized triangular (3-coordinated) and hexagonal (6-coordinated) holes have well-defined radii of $r/\mathcal{D} = 0.6435$ and 1.115, respectively, which are the radii of the circumscribed circles of the polygons formed by the disk centers.
The 9- and 12-coordinated holes have two characteristic length scales, given by the radii of the inscribe and the circumscribed circles.
We find that the radii of the circumscribed circles of the 3-, 6-, 9-coordinated idealized holes provide good lower bounds of the first three maxima in $G_V(r)$, respectively; see Fig. \ref{fig:2DHSphi0x73}(a).
These lower bounds are expected, since to create holes slightly larger than the circumscribed circles of the idealized holes, all particles surrounding the hole must be simultaneously displaced away from the hole center.
Such significant distortions of the crystal geometry result in maxima in $G_V(r)$.
While we do not report data for $G_V(r)$ beyond 2$\mathcal{D}$, we expect another local maximum of $G_V(r)$ to occur just above the radius of the circumscribed circle of the idealized 12-coordinated hole at $r = 2.229\mathcal{D}$.
On the other hand, the local minima in $G_V(r)$, which indicate stable circular-like hole sizes, occur at $r$ values smaller than the radii for the corresponding idealized holes in Fig. \ref{fig:hole_idealized}.\footnote{For 9- and 12-coordinated holes, the hole radius is taken to be the average of the radii of the inscribed and the circumscribed circles.}
This observation is consistent with the well-known effect that a vacancy in an equilibrium crystal has asymmetric fluctuations in the radius of the largest spherical hole that it can contain. 
The circular-like vacancies tend to compress rather than expand compared to their sizes in the perfect crystal \cite{Ba00, Pr01}.
Compression of the vacancies increases the free volumes of the remaining particles, thereby increasing the entropy of the crystal.
Furthermore, as particles surrounding the vacancies undergo asymmetric displacements, the vacancies become aspherical, which again decreases the radius of the largest spherical hole contained in the vacancy.

\begin{figure}[htp]
    \centering
    \includegraphics[width=110mm]{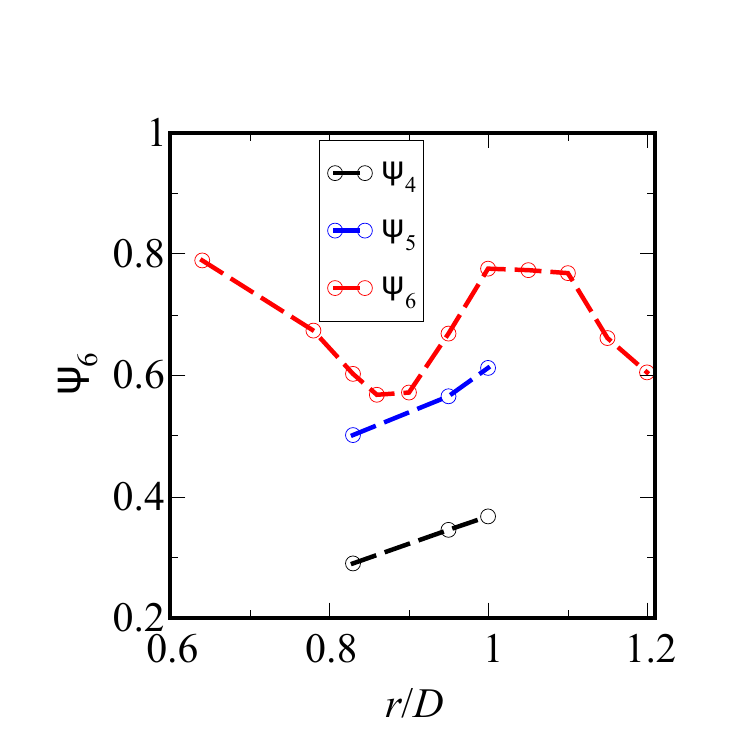}
    \caption{
    Local orientational order metrics $\psi_4$, $\psi_5$ and $\psi_6$ (\ref{psin}) for the nearest 18 particles to the test point in a 2D hard disk crystal at $\phi = 0.73$.}
    \label{fig:2DHSphi0x73_Phi6}
\end{figure}

We expect that the effect of holes on the local crystalline configuration to be reflected in the local orientational order in the vicinity of the holes.
To study the correlation between the oscillations in $G_V(r)$ and local orientational order, we compute the bond-orientational order parameter $\psi_n$ \cite{Lev86}, which measures the $n$-fold orientational order in a 2D many-body system, given by
\begin{equation}
    \psi_n = \left|\frac{1}{M}\sum_{j = 1}^M \exp(i n\theta_{j})\right|,
    \label{psin}
\end{equation}
where the sum is over $M$ nearest neighbors $\mathbf{r}_j$ of the test particle at $\mathbf{t}$, $\theta_j$ is the angle between the vector $\mathbf{r}_j - \mathbf{t}$ and an arbitrary but fixed reference vector \footnote{The special case of $\psi_6$ was first proposed by Nelson and Halperin} \cite{Ne79}. 
In this work, we take $M = 18$, a choice that considers hard disks in the first 2--3 coordination shells of the test point.
The red curve in Fig. \ref{fig:2DHSphi0x73_Phi6} plots $\psi_6$ for the 2D hard-disk crystal at $\phi = 0.73$ as a function of the hole radius $r$. 
It is clear that $\psi_6$ in the vicinity of the test point exhibits oscillating behaviors as $r$ increases, with a minimum at $r = 0.86\mathcal{D}$ and a maximum at $r = \mathcal{D}$.
This shows that the hexatic orientational order of the triangle lattice is preserved with triangular holes and hexagonal holes, but is disrupted at hole radii that lie between these two ``stable'' local hole geometries.
Figure \ref{fig:2DHSphi0x73_Phi6} also shows $\psi_4$ and $\psi_5$ with $M = 18$ at hole sizes near $r = 0.86\mathcal{D}$. 
We observe that while $\psi_6$ has a local minimum for these hole sizes, it is still larger than $\psi_4$ and $\psi_5$. 
Thus, while these hole sizes may be 4- or 5-coordinated [e.g., as shown in Fig. \ref{fig:hole_geom}(b),] the second and third coordination shells still possess hexatic orientational order to a larger extent than 4- or 5-fold orientational order.

\begin{figure}[htp]
    \centering
    \subfloat[]{\includegraphics[width=80mm]{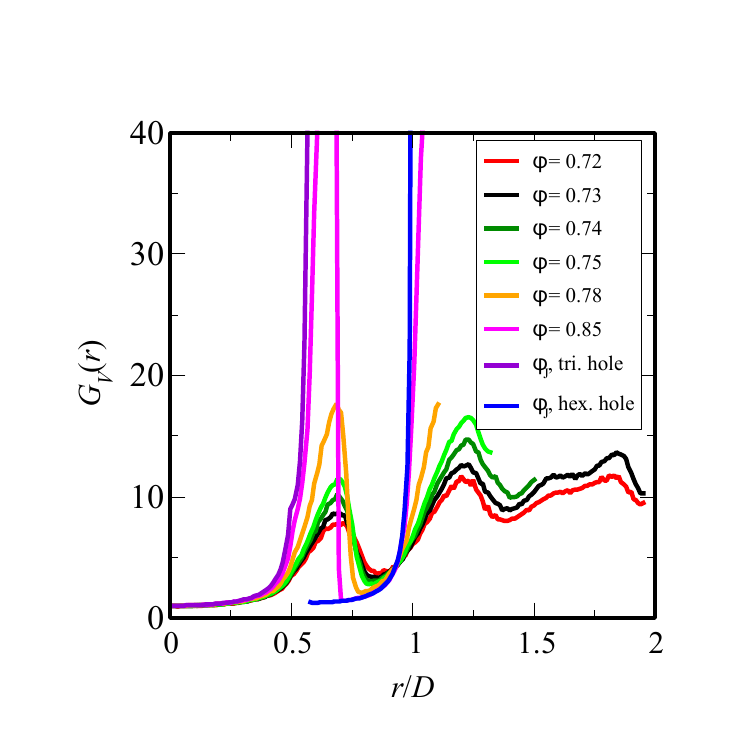}}
    \subfloat[]{\includegraphics[width=80mm]{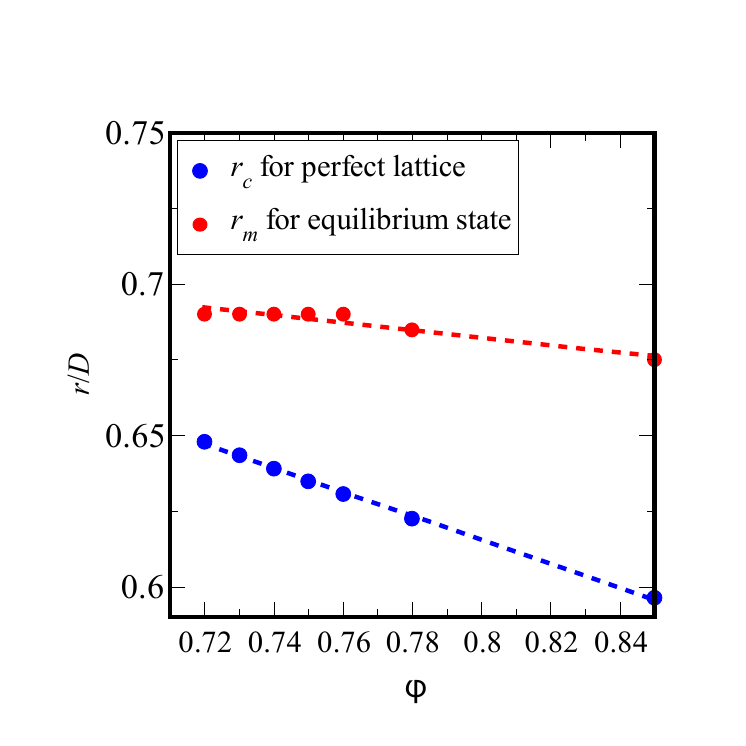}}
    \caption{
    (a) The function $G_V(r)$ for 2D hard-disk crystals at different packing fractions obtained from the grand-canonical biased-sampling scheme.
    (b) Plots of the covering radius $r_c(\phi)$ of the perfect triangle lattice and the position of the first peak in $G_V(r)$ for equilibrium hard-disk crystals $r_m(\phi)$.}
    \label{fig:2DHSdifferentPhi}
\end{figure}

Figure \ref{fig:2DHSdifferentPhi}(a) shows $G_V(r)$ for 2D crystals at various packing fractions.
As $\phi$ increases, the amplitudes of the oscillations grow rapidly and the first maximum of $G_V(r)$ occurs at decreasing $r$ values. 
As $\phi$ approaches the jamming packing fraction $\phi_J = \pi/\sqrt{12}$ for the triangle lattice, the position of the first peak approaches the so-called covering radius $r_c(\phi_J) = \mathcal{D}/\sqrt{3} = 0.577\mathcal{D}$ \cite{To10d}, which is the distance from the centroid of the triangular hole to the center of any hard disk on the vertex of the triangular hole.
It is known that at $\phi = \phi_J$, $G_V(r)$ develops a pole of order one at $r_c(\phi_J)$ \cite{Mi20}.
At all 2D packing fractions that we study, we find that the covering radius $r_c(\phi)$ provides a lower bound of the position of the first peak $r_m$ in $G_V(r)$, as shown in Fig. \ref{fig:2DHSdifferentPhi}(b).
In addition, as $\phi \rightarrow \phi_J$, the second peak of $G_V(r)$ approaches $G_V(r)$ for a hexagonal hole created by removing one particle in the prefect triangle lattice [dark blue curve in Fig. \ref{fig:2DHSdifferentPhi}(a)], the expression of which is the same as $G_V(r)$ for the densest honeycomb packing of hard disks on the range $\mathcal{D}/\sqrt{3} \leq r\leq \mathcal{D}$, given in Ref. \citenum{To10d}.

\subsubsection{Dense Fluid and Hexatic States}
\begin{figure}[]
    \centering
    \subfloat[]{\includegraphics[width=80mm]{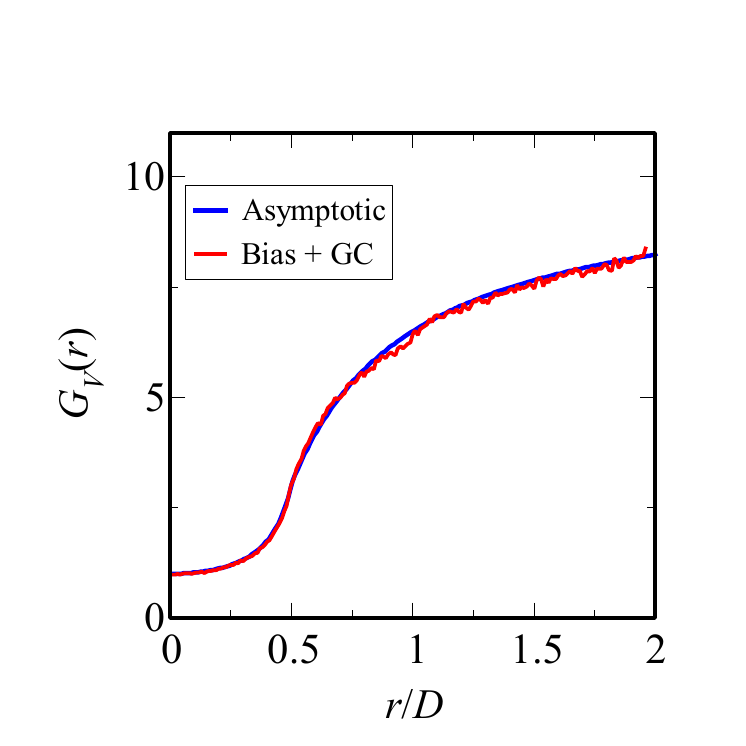}}
    \subfloat[]{\includegraphics[width=80mm]{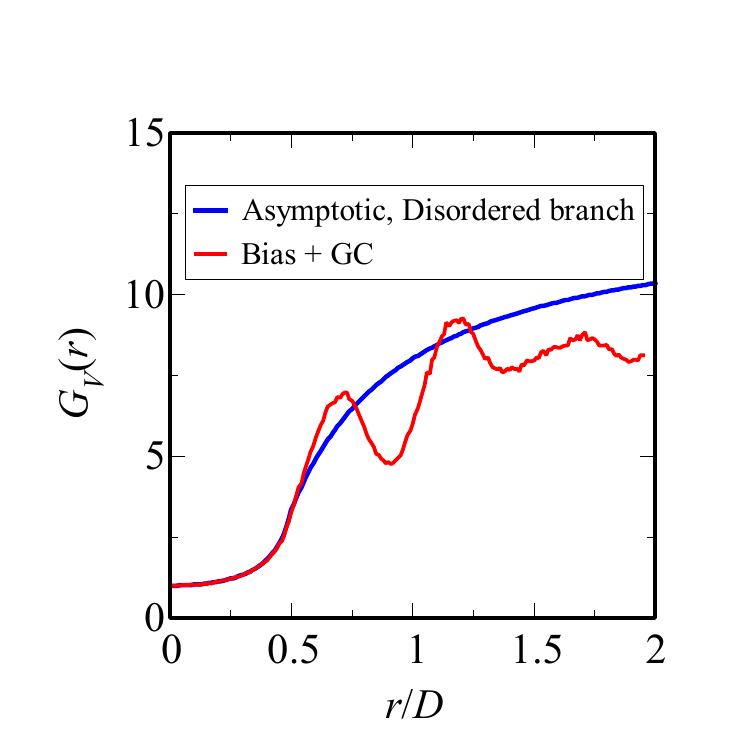}}
    \caption{(a) The function $G_V(r)$ for 2D hard-disk fluid at $\phi = 0.670$ obtained from the grand-canonical biased-sampling scheme, as well as the asymptotic expression via Eqs. (\ref{gv2d}) and (\ref{a0a12d}). 
    (b) The function $G_V(r)$ for 2D hard-disk hexatic state at $\phi = 0.702$ obtained from the grand-canonical biased-sampling scheme, as well as asymptotic expression extrapolated from the disordered branch [Eqs. (\ref{gv2d}) and  (\ref{a0a12d})].}
    \label{fig:2DHSphi0x670x70}
\end{figure}

Because oscillations in $G_V(r)$ are already significant for 2D crystals at packing fractions well below jamming, we investigate whether they also appear at lower packing fractions, i.e., in dense fluid and hexatic phases.
Figure \ref{fig:2DHSphi0x670x70}(a) and (b) shows $G_V(r)$ for the 2D equilibrium hard-disk system in the dense fluid phase at $\phi = 0.670$ and the hexatic phase at $\phi = 0.702$, respectively. 
Note that in the former case, $\phi$ is just below the first-order fluid-hexatic transition packing fraction $\phi_f = 0.69$ \cite{Ho68, Ko08}.
Figure \ref{fig:2DHSphi0x670x70}(a) also plots the asymptotic expression for $G_V(r)$ along the disordered branch predicted from Eqs. (\ref{gv2d}) and (\ref{a0a12d}).
It is obvious the simulated $G_V(r)$ for the fluid state is monotonically increasing with $r$ and agrees closely with the asymptotic expression. 
We remark that the monotonicity of $G_V(r)$ is in strong contrast with the functional form of the radial distribution functions $g_2(r)$ for dense hard-disk fluids, which possess strong oscillations up to $r = 4\mathcal{D}$ \cite{Mo05}.
On the other hand, Fig. \ref{fig:2DHSphi0x670x70}(b) shows that $G_V(r)$ for the hexatic state is no longer monotonic in $r$.
The existence of oscillations in $G_V(r)$ is consistent with the statement that the hexatic phase is orientationally ordered.
It is also noteworthy that for most values of $r$ on the interval $(\mathcal{D}, 1.95\mathcal{D}$), the simulated $G_V(r)$ at large $r$ is smaller than the asymptotic expressions extrapolated from the disordered branch, Eqs. (\ref{gv2d}) and  (\ref{a0a12d}).
The simulated value of $G_V(\mathcal{D})$ is also smaller than the value predicted using Eqs. (\ref{gv2d}) and  (\ref{a0a12d}).
Using Eq. (\ref{pressure}), we deduce that the pressure of the hexatic phase is lower than the disordered metastable state at the same packing fraction.

\subsection{3D FCC Hard-Sphere Crystals}
\begin{figure}[htp]
    \centering
    \subfloat[]{\includegraphics[width=80mm]{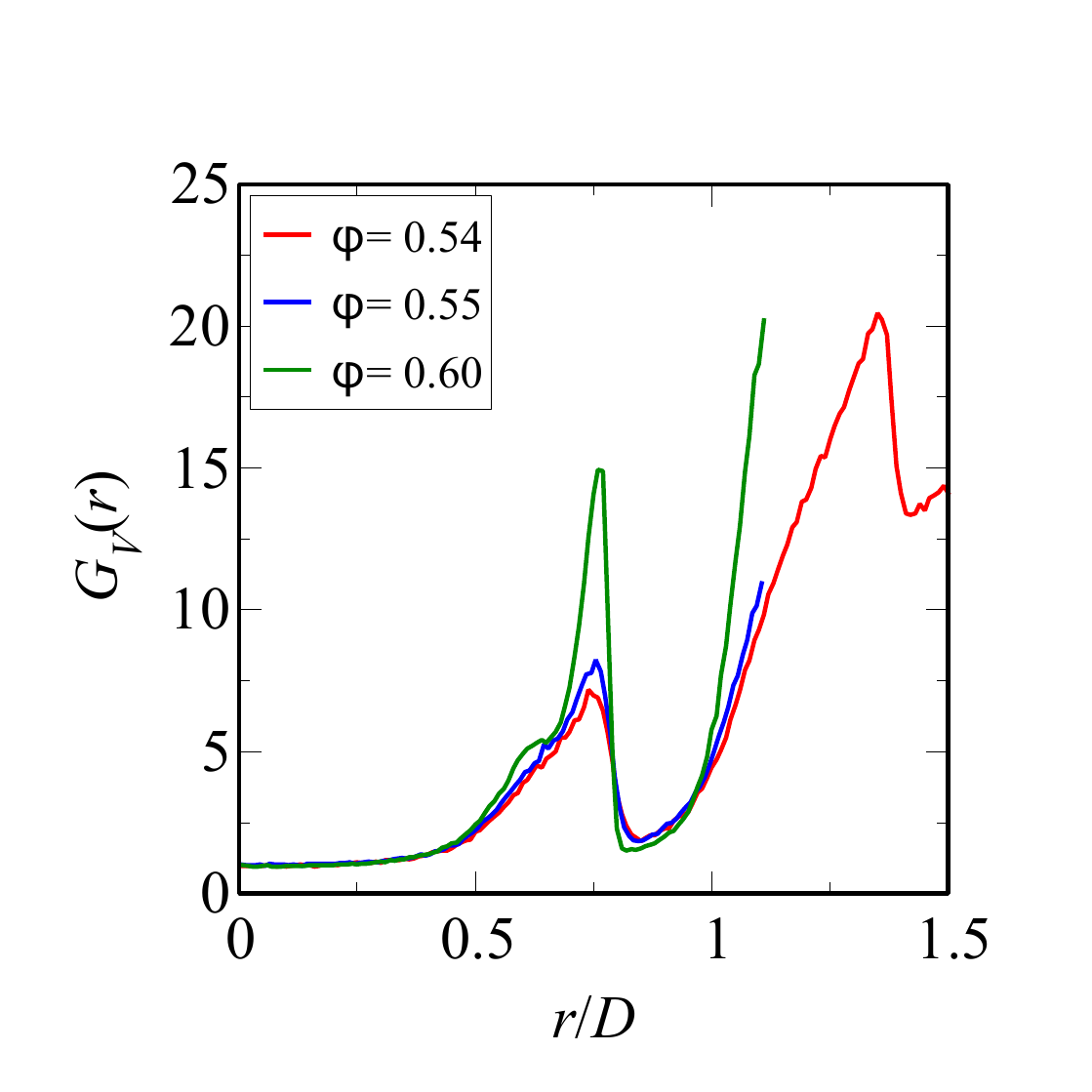}}
    \subfloat[]{\includegraphics[width=80mm]{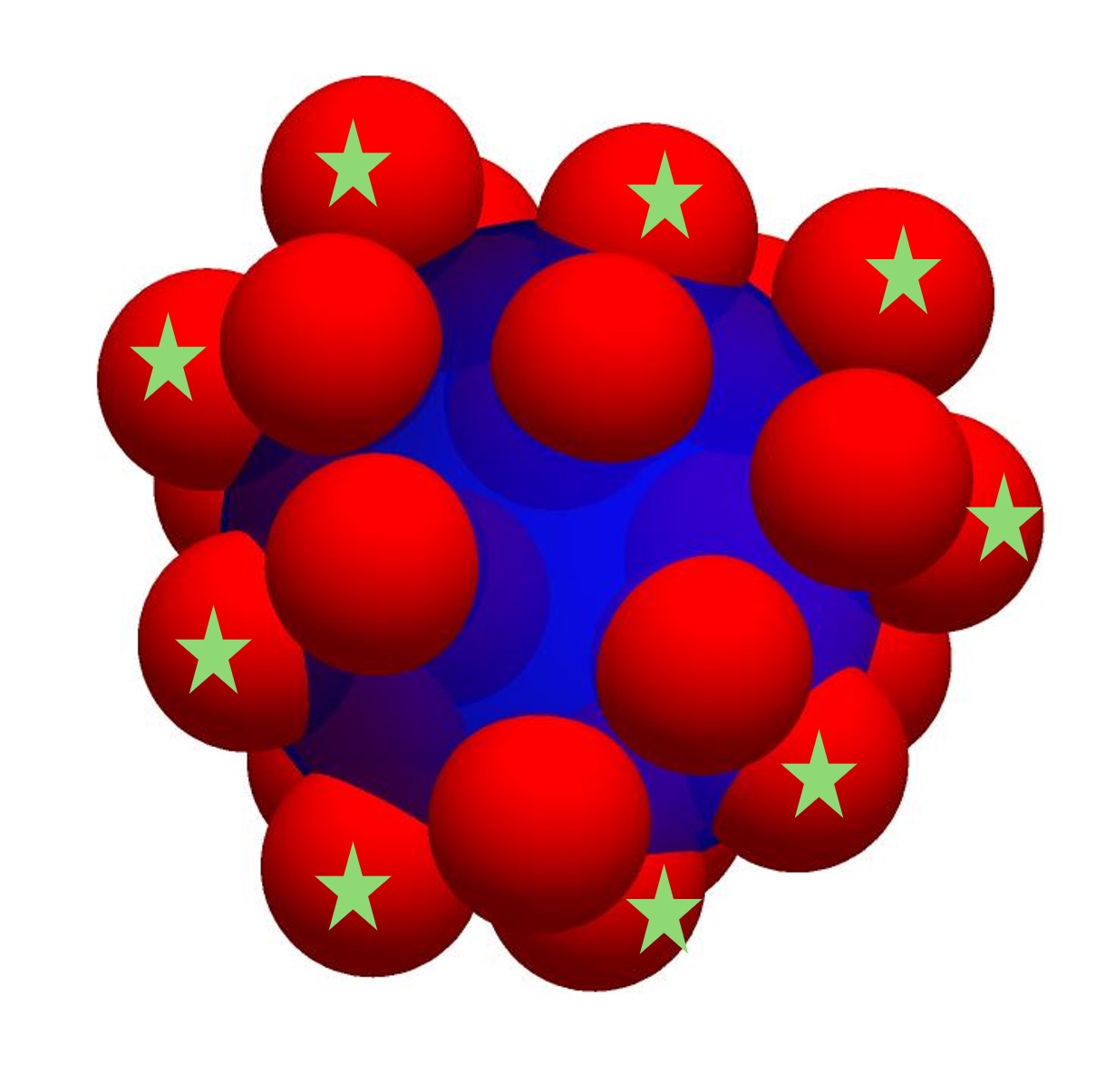}}
    \caption{(a) The function $G_V(r)$ for the 3D hard-sphere systems at $\phi = 0.54, 0.55$ and $0.60$ obtained via biased sampling in the grand canonical ensemble.
    (b) Local coordination configuration of a hole (indicated with a blue sphere) in an equilibrium hard-sphere system at $\phi = 0.54$ of radius $r = 1.47\mathcal{D}$, approximately at the second minimum in $G_V(r)$.
    The 9 hard-sphere particles along a great circle of this hole are indicated with stars.
    }
    \label{fig:3DHS}
\end{figure}

Figure \ref{fig:3DHS}(a) show $G_V(r)$ for the equilibrium 3D hard-sphere fcc-lattice crystal at three different packing fractions. 
The cases $\phi = 0.55$ and 0.60 are slightly above the melting packing fraction $\phi_m = 0.545$ \cite{Ho68}. 
The case $\phi = 0.54$, slightly below the melting point, is chosen to facilitate an efficient and accurate calculation of $G_V(r)$ up to $r = 1.5\mathcal{D}$, thereby providing preliminary qualitative information about hole statistics for 3D crystals near melting.
For $\phi = 0.54$, the proportion of the liquid phase is expected to be negligible with $\langle N\rangle = 256$ particles used in our study, since $\phi$ is very close to $\phi_m$. 
\footnote{It would be useful for a future study to examine substantially larger systems to test this expectation.}
Indeed, we have not observed phase separation at $\phi = 0.54$ in our simulations.

As is the case for the 2D hard-disk crystal, we find significant oscillations in $G_V(r)$ with amplitudes that rapidly increase with $\phi$, and the period of the oscillations is approximately $\mathcal{D}/2$.
The first minimum of $G_V(r)$ corresponds to a vacancy created by the removal of one particle in the crystal.
The coordination number of such holes is $Z = 12$.
Figure \ref{fig:3DHS}(b) shows a representative local coordination configuration of a hole of radius $r = 1.47\mathcal{D}$ in the hard-sphere system at $\phi = 0.54$, i.e., approximately at the second minimum in $G_V(r)$ at $r = 1.42\mathcal{D}$.
Such holes have $Z = 24$ and correspond to cavities in the fcc crystal created by removing 4 particles that form the vertices of a tetrahedral hole.
Note that in a perfect close-packed fcc lattice, there are 25 particles that touch a tetrahedral 4-particle cluster. 
In equilibrium hard-sphere crystals, $Z$ is slightly lower than 25 due to the fact that the 25-coordinated cavity created from the perfect fcc is highly aspherical, and they are distorted in the equilibrium states to yield nearly spherical holes.
Specifically, we find that there are usually 9 hard-sphere particles fitted along a great circle of these holes, indicated with stars in Fig. \ref{fig:3DHS}(b).
These 9-coordinated big circles are analogous to the 2D local coordination configuration shown in Fig. \ref{fig:hole_geom}(e).
Note that the second maximum and minimum in $G_V(r)$, as well as the tetrahedron-like holes, are features that are almost certainly qualitatively correct for the crystal phase near melting.
Indeed, the tetrahedron-like holes have also been observed in simulations at $\phi = 0.55$ and 0.60. 
However, we do not include statistics for these holes in Fig. \ref{fig:3DHS}(a), because $G_V(r)$  near $r = 1.5\mathcal{D}$ did not converge within $2\times 10^6$ sweeps at these packing fractions.

In contrast to $G_V(r)$ for 2D hard-disk crystals [Fig. \ref{fig:2DHSphi0x73}(a) and (c)], we observe in Fig. \ref{fig:3DHS} that for the fcc crystals, the first peak of $G_V(r)$ has a shoulder at $r \sim 0.65\mathcal{D}$, which becomes more apparent as $\phi$ increases.
This is due to the fact that the 2D triangle-lattice packing contains only triangular holes, whereas the 3D fcc packing contains both tetrahedral and octahedral holes. 
In a perfect fcc lattice with packing fraction $\phi$, the radius of the tetrahedral holes is given by 
\begin{equation}
    r_t(\phi) = \sqrt{\frac{3}{8}}\left(\frac{\phi_J}{\phi}\right)^{1/3}\mathcal{D},
\end{equation}
where $\phi_J = \pi/\sqrt{18}$ is the packing fraction for the fcc close packing.
Thus, we have $r_t(0.55) = 0.676\mathcal{D}$ and $r_t(0.60) = 0.657\mathcal{D}$, which are approximately the hole radii where the shoulders in  $G_V(r)$ appear in Fig. \ref{fig:3DHS}.
Thus, the shoulders represent the transition between smaller holes that may be derived from either tetrahedral or octahedral holes in the fcc lattice and larger holes that are exclusively derived from octahedral holes.
Analogous to the 2D case, $r_t(\phi)$ provides a lower bound of the position of the first peak in $G_V(r)$.

\section{Local Packing Fraction Distribution of Delaunay Cells}
\label{sec_localphi}
    \begin{figure}[htp]
    \subfloat[]{\includegraphics[width=80mm]{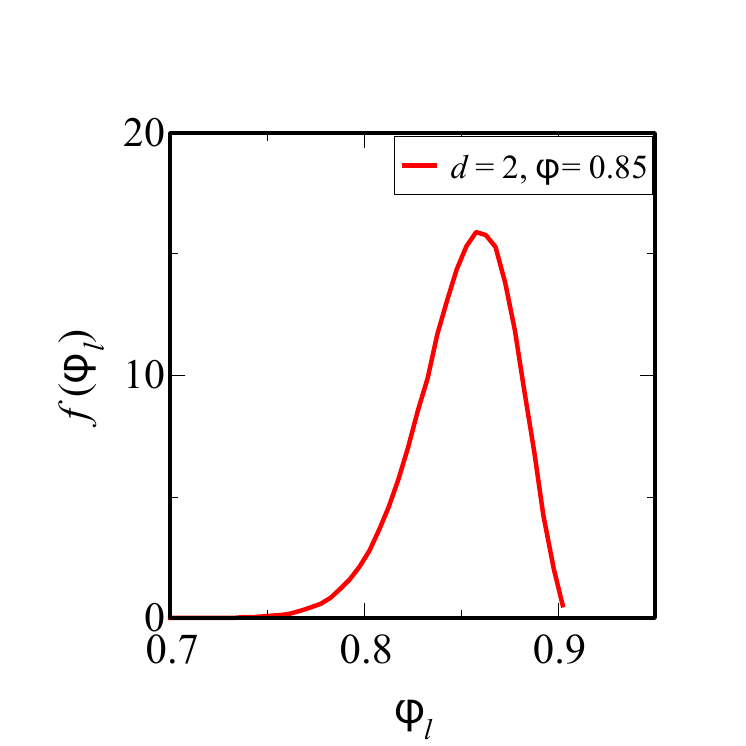}}
    \subfloat[]{\includegraphics[width=80mm]{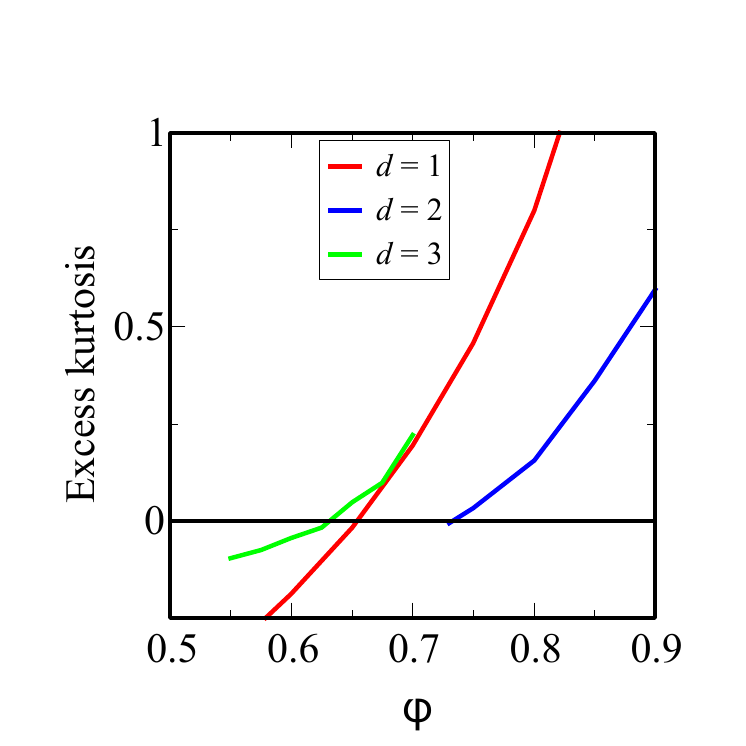}}
    \caption{(a) The probability distribution $f(\phi_l)$ of the local packing fraction of Delaunay cells in the 2D hard-disk crystal at $\phi = 0.85$. 
    (b) The excess kurtosis of $f(\phi_l)$ for high-density hard-sphere systems as a function of $\phi$ in one, two and three dimensions.}
    \label{localphi}
\end{figure}

Here, to further characterize the distribution of interparticle spacing in configurations of hard-sphere systems, we study the local packing fraction distribution of Delaunay cells as a function of the global packing fraction $\phi$ in one, two and three dimensions.  
For a point configuration $\mathbf{r}^N$ in $\mathbb{R}^d$, a Delaunay tessellation is a $d$-dimensional triangulation $\operatorname{DT}(\mathbf{r}^N)$ such that no point in $\mathbf{r}^N$ is inside the circum-hypersphere of any $d$-simplex in $\operatorname{DT}(\mathbf{r}^N)$ \cite{Wa81}.
Here, we use point configurations derived from the sphere centers in equilibrium $d$-dimensional hard-sphere systems in the canonical ensemble with $N \sim 500$ particles (without the test point) to create the corresponding Delaunay tessellations.
We then compute the local packing fraction $\phi_l$, i.e., the fraction of the particle phase in each Delaunay cell, and determine the distribution of $\phi_l$ over all Delaunay cells in 200 independent configurations, denoted by $f(\phi_l)$.
Figure \ref{localphi}(a) shows $f(\phi_l)$ for the equilibrium 2D hard-disk crystal at global packing fraction $\phi = 0.85$.
Apparently, $f(\phi_l)$ is asymmetric about $\phi$, as it resembles a Gaussian for $\phi_l < \phi$ but decays more rapidly for $\phi_l > \phi$. 
This is due to the fact that $\phi = 0.85$ is close to the close packing fraction $\phi_f = \pi/\sqrt{12} =0.9069$, which is an upper bound of $\phi_l$.

To more closely investigate the non-Gaussian behaviors of the local packing fraction distribution across $\phi$ and $d$, we plot in Fig. \ref{localphi}(b) the excess kurtosis of $f(\phi_l)$ for equilibrium 1D high-density hard-rod fluids as well as for equilibrium 2D and 3D hard-sphere crystals.
The kurtosis of a random variable $X$ with mean $\mu$ and standard deviation $\sigma$ is given by
\begin{equation}
    \operatorname{Kurt}(X) = \operatorname{E}\left[\left(\frac{X - \mu}{\sigma}\right)^4\right],
\end{equation}
and the excess kurtosis is defined as $\operatorname{Kurt}(X) - 3$ \cite{Ba88}.
A distribution with negative excess kurtosis is known as platykurtic, indicating that it contains fewer or less extreme outliers than the normal distribution of the same mean and standard deviation \cite{Ba88}.
Interestingly, Fig. \ref{localphi}(b) shows that in each of the first three dimensions, the excess kurtosis increases with $\phi$ and switches from negative to positive at a certain dimension-dependent transition packing fraction $\phi^*$.
Indeed, we observe that for $\phi$ away from jamming, where the excess kurtosis is negative, the shape of $f(\phi_l)$ is approximately symmetric and Gaussian-like. 
Thus, it is expected that the distribution is platykurtic at low $\phi$, because $f(\phi_l)$ has a finite support $[0, \phi_J]$, and its tails on both ends must decay faster than a Gaussian.
However, at high $\phi$ close to jamming, $f(\phi_l)$ is asymmetric and has a heavier tail toward smaller $\phi_l$, indicating that the hard spheres undergo collective motion to creating regions of significantly lower local packing fractions than $\phi$.

For the 2D state, $\phi^* = 0.73$ is slightly above the packing fraction for the hexatic-crystal transition.
For the 3D state, $\phi^* = 0.625$ lies between the melting packing fraction and the fcc close-packing fraction.
Furthermore, we find that as the dimensionality increases, the magnitudes of the excess kurtosis are closer to zero, i.e., $f(\phi_l)$ becomes closer to the Gaussian distribution.
The variation of the excess kurtosis with $\phi$ is also less steep in higher dimensions.
This reflects the decorrelation principle, which states that unconstrained spatial correlations vanish asymptotically for pair distances beyond the hard-core diameter in the high-dimensional limit \cite{To06b}.
As the particles become increasingly decorrelated in higher dimensions, $f(\phi_l)$ is expected to approach a Gaussian as a result of the central limit theorem.

\section{Discussion and Conclusions}
\label{conc}
In summary, we have introduced a biased-sampling scheme that accurately determine void nearest-neighbor functions, i.e., the hole statistics, for dense hard-sphere crystals and liquids on ranges of the hole radius $r$ that far extends the sizes that could be previously explored.
Using this algorithm, we find that for 2D hexatic and crystal phases, $G_V(r)$ exhibits oscillations with amplitudes that rapidly increase with the packing fraction.
Such oscillations stand in contrast to the 2D fluid states, for which $G_V(r)$ is monotonic up to the fluid-hexatic transition point.
For 3D crystals, $G_V(r)$ is also non-monotonic, and a shoulder develops its first peak as $\phi$ increases,  indicating the transition between tetrahedral to octahedral holes in the fcc crystal.
We find that the oscillations in $G_V(r)$ for order states are strongly correlated with the local orientation order.
For both 2D and 3D systems, minima of $G_V(r)$ correspond to stable hole whose local coordination geometry is compatible with the crystal structure, whereas the maxima of $G_V(r)$ correspond to geometrically strained holes that significantly disturb the local crystal structure.
We have also studied the local packing fraction distribution $f(\phi_l)$ of Delaunay cells and find that the excess kurtosis of $f(\phi_l)$ increases with increasing $\phi$, and changes sign at a dimension-dependent transition packing fraction $\phi^*$.
This switch from a platykurtic to a leptokurtic behaviors in $f(\phi_l)$ as $\phi$ increases indicates the transition between predominantly weakly correlated displacements of hard spheres at lower $\phi$ to significant collective displacements at higher $\phi$.
It is likely that $\phi^*$ lies in the crystal phase also for certain higher dimensions $d \geq 4$.

It is important to remark on the relationship between the void nearest-neighbor functions and the $n$-body correlation function $g_n$ for hard-sphere systems, which has been originally explored by Reiss et al \cite{Re59}.
The exact series representation of $G_V(r)$ in terms of $g_n$ is given by \cite{To90c}:
\begin{equation}
    G_V(r) = \lim_{|\textbf{x} - \textbf{r}_1| \rightarrow r} \frac{1}{\rho E_V(r)} \left[\rho + \sum_{k = 1}^\infty \frac{(-1)^k\rho^{k+1}}{k!} \int g_{k + 1}(\mathbf{r}^{k + 1})\prod_{i = 2}^{k + 1}m(|\mathbf{x} - \mathbf{r}_i|; r) d\mathbf{r}_i\right].
    \label{Gv-g}
\end{equation}
where 
\begin{equation}
    m(y; r) = 
    \begin{cases}
        1, \quad y < r\\
        0, \quad y \geq r,
    \end{cases}
\end{equation}
and the series representation of $E_V(r)$ is given by
\begin{equation}
    E_V(r) = 1 + \Sigma_{k = 1}^\infty (-1)^k\frac{\rho^k}{k!}\int g_k(\mathbf{r}_1, \dots \mathbf{r}_k)v_k^{int}(\mathbf{r}_1, \dots \mathbf{r}_k; r)d\mathbf{r}_1\dots d\mathbf{r}_k,
\end{equation}
where $v_k^{int}(\mathbf{r}_1, \dots \mathbf{r}_k; r)$ is the intersection volume of $k$ equal spheres of radius $r$ centered at positions $\mathbf{r}_1, \dots \mathbf{r}_k$.
Because $G_V(r)$ involves an integral of $g_2(r)$ in Eq. (\ref{Gv-g}), any singularity in $g_2(r)$ translates to a corresponding singularity in $G_V(r)$ of a higher order.
Indeed, Reiss et al. \cite{Re59} explicitly demonstrated that $G_V(r)$ is smoother than $g_2(r)$ on certain ranges, because $g_2(r)$ on $[\mathcal{D}, 2\mathcal{D}/\sqrt{3}]$ can be exactly expressed in terms of the first two derivatives of $G_V(r)$ on $[\mathcal{D}/2, \mathcal{D}/\sqrt{3}]$.
These authors also show that there exist singularities in $G_V(r)$ at $r = \mathcal{D}/2, \mathcal{D}/\sqrt{3}, \dots$, i.e., the hole radii at which $g_2, g_3, \dots$ just become non-vanishing in (\ref{Gv-g}) \cite{Re59}.
Moreover, in light of the fact that Stillinger \cite{St71} has shown that $g_2(r)$ for 2D and 3D hard spheres contain singularities for all $r > \mathcal{D}$ at positive number density \cite{Note6}, $G_V(r)$ must also be nonanalytic for all $r\geq \mathcal{D}/2$, with dense singularities of higher orders than the corresponding ones in $g_2(r)$ \cite{Note7}.
\footnotetext[6]{Here, the condition of strictly positive density is important to show the dense singularities in $g_2(r)$, because the virial expansion does not show singularities for all $r$.}
\footnotetext[7]{We remark that Reiss et al. \cite{Re59} were not aware of the result in Ref. \citenum{St71} that $g_2(r)$ is nonanalytic at all $r > \mathcal{D}$.}

\begin{figure}[htp]
    \subfloat[]{\includegraphics[width=78mm, trim = {0, -0.55cm, 0, 0}, clip]{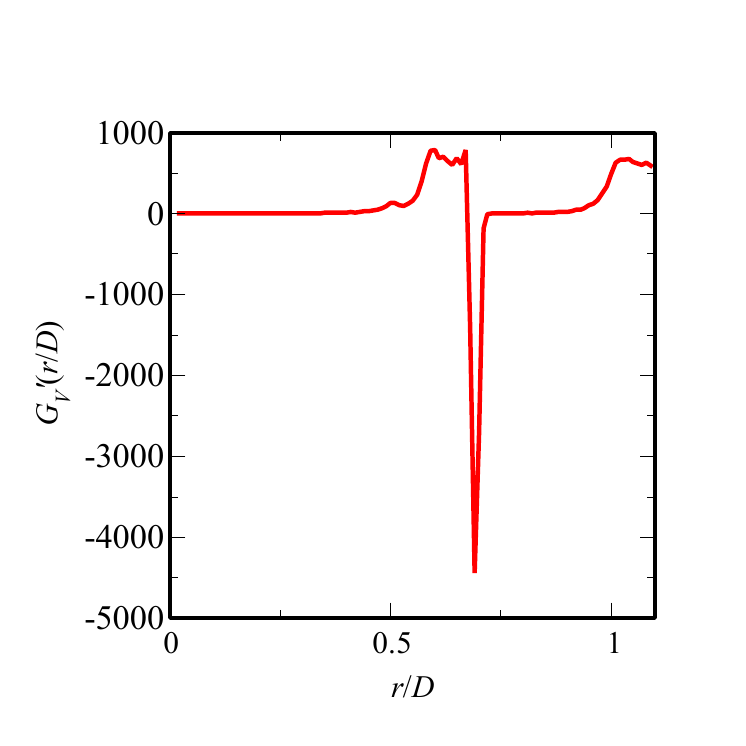}}
    \subfloat[]{\includegraphics[width=82mm]{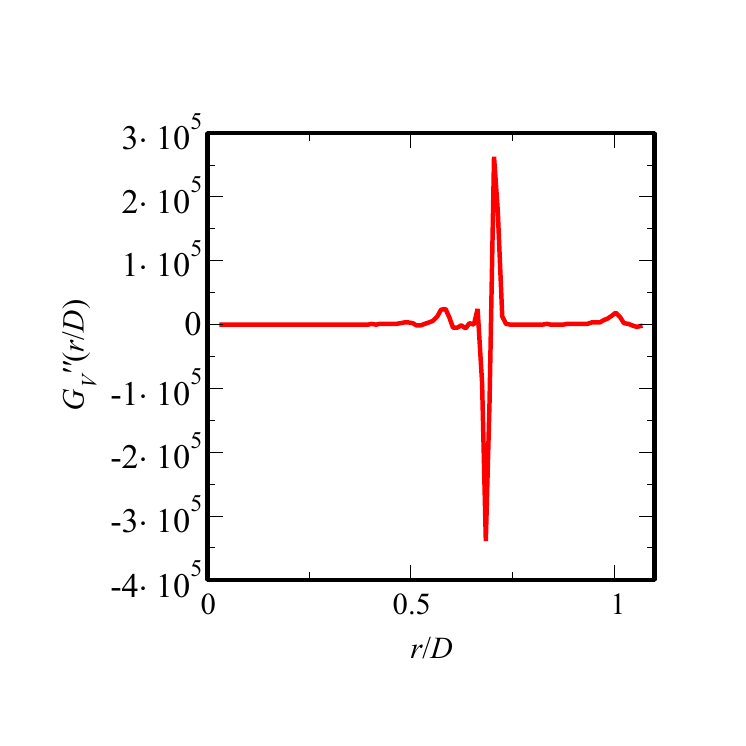}}
    \caption{(a) First and (b) second derivatives   $G_V'(r/\mathcal{D})$ and $G_V''(r/\mathcal{D})$, respectively, of $G_V(r/\mathcal{D})$ for the 2D hard-disk crystal at $\phi = 0.85$. 
    The first derivative increases steeply at $r/\mathcal{D} = 0.58$ and 1.0, and plateaus in the ranges $[0.60, 0.67]$ and $[1, 1.2]$.
    Spikes with very large magnitudes are observed in both derivatives, indicating that $G_V(r)$ is nearly discontinuous at $r\sim 0.70\mathcal{D}$.
    Except for the spike, the second derivative ([curvature of $G_V(r)$]) is relatively flat.}
    \label{fig:derivatives}
\end{figure}

Figure \ref{fig:derivatives}(a) and (b) show the first and second derivatives $G_V'(r/\mathcal{D})$ and $G_V''(r/\mathcal{D})$, respectively, of $G_V(r/\mathcal{D})$ for the 2D hard-disk crystal at $\phi = 0.85$. 
The first derivative $G_V'(r/\mathcal{D})$ drops sharply to $-4\times 10^3$ $r = 0.70\mathcal{D}$.
The magnitudes of the second derivative in the vicinity of this radius reach as large as $3\times 10^5$.
Thus, for hard-sphere packings close to jamming, $G_V(r)$ can be considered to be nearly discontinuous in the first and higher derivatives. 
As $\phi$ approaches $\phi_J$, $G_V(r)$ becomes truly discontinuous at the covering radius for the close-packing triangle lattice $r_c(\phi_J) =  0.577\mathcal{D}$.

Apart from the spike, we find that $G_V'(r/\mathcal{D})$ steeply increases at $r/\mathcal{D} = 0.58$ and has a plateau on the range $[0.60, 0.67]$.
Small holes of $r \leq 0.58\mathcal{D}$ can be fitted into the triangular holes of the 2D crystal without disturbing the local hexatic symmetry. Thus, increasing $r$ implies that the test point must be placed closer to the centroids of the triangular holes.
The increase of $G_V'(r/\mathcal{D})$ due to the effect of test-point positions can be predicted from the expression of $G_V(r)$ for the perfect triangle lattice \cite{To10d}, which exhibits rapid increases in all derivatives in $G_V(r)$ up to the singularity at the covering radius.
By contrast, for the range $[0.60, 0.67]$, the key contribution to variations in $G_V(r)$ is the expansion of the triangular holes.
Similarly, the increase of $G_V'(r/\mathcal{D})$ at $r = \mathcal{D}$ and the plateau beyond this radius are due to the position of the test point toward the centroids of the hexagonal holes and the expansion of the hexagonal holes, respectively.
We further remark that except for the spikes, $G_V'(r/\mathcal{D})$ exhibits less significant oscillations than $G_V(r)$ itself, and the second derivative $G_V''(r/\mathcal{D})$ [curvature of $G_V(r)$] is rather constant in $r$.
These behaviors suggest that the discontinuities in the first and second derivatives of $G_V(r)$ are primarily due to the abrupt transition between triangular and hexatic stable holes at $r \sim 0.70\mathcal{D}$, and $G_V(r)$ values away from this transitional radius  can be well approximated by polynomials or rational functions.

Our investigation of hard spheres facilitates the study of structural and bulk properties of realistic materials.
While in actual crystalline materials, the interactions between the constituent particles are different from hard-sphere interactions, we expect similar oscillatory behavior of $G_V(r)$ to occur, since the hard-sphere system is a general idealized model for systems with strong short-ranged repulsive interactions.
Furthermore, our investigation of the local density distribution can also be generalized to systems with non-hard-core interactions, such as the well-known Lennard-Jones or Gaussian-core models \cite{St81}. 
For such systems, $\phi_l$ is not well-defined, but one can study the distribution of the inverse volumes of the Delaunay cells, and we expect that the kurtosis of this distribution to increase with the global number density $\rho$.
Finally, we note that because a hole in a many-body system can be considered as a ``solute'' particle in a ``solvent'' of particles \cite{Re59}, information about the relative stability of hole sizes in crystals, as manifested by the oscillations in $G_V(r)$, could enable the prediction of solubility in alloys and other solid solutions \cite{Ax48, So10}.

While the radial void nearest-neighbor functions [such as $G_V(r)$] are essential in the study of thermodynamic properties of hard-sphere systems (see Sec. \ref{def}), they do not explicitly reveal detailed statistics for aspherical holes.
In equilibrium crystals, it is possible to create aspherical vacancies.
For example, removing two neighboring particles in 2D and 3D crystals generates an elongated vacancy, whereas removing 3 neighboring particles in a 3D crystal generates a flat vacancy. 
In such cases, $G_V(r)$ samples spherical void regions contained within such vacancies, and information about the shape of larger aspherical vacancies may be hidden in its higher derivatives.
However, one could probe the statistics of aspherical holes by considering the analogous vector-dependent function $G_V(\mathbf{r})$ which is proportional to the conditional probability of finding a particle at $\mathbf{r}$ from the test point, given that no particle center is found on the ray $\{\lambda\mathbf{r}: \lambda \in [0, 1)\}$.
For example, in two dimensions, one can define $G_V(\mathbf{r}) = G_V(r, \theta)$ such that $\rho G_V(r, \theta)rdrd\theta$ is the conditional probability of finding a particle center at polar coordinates $(r, \theta)$, given that no particle center is found in the sector defined by the region $\{(r',\theta'): r'\in[0, r), \theta'\in(\theta - d\theta/2, \theta + d\theta/2)\}$.
Precise determination of the vector-dependent hole statistics can enable the study of solubility in molecular solid solutions in which the solutes are typically aspherical molecules \cite{Lu18}.

\clearpage
%

\end{document}